# Some Dynamic Market Models

*Jan A. Audestad*

Norwegian University of Science and Technology,
Gjøvik University College

**ABSTRACT**

In this text, we study the behavior of markets using models expressible as ordinary differential equations. The markets studied are those where each customer buys only one copy of the good, for example, subscription of smartphone service, journals and newspapers, and goods such as books, music and games.

One of the underlying models is the Bass model for market evolution. This model contains two types of customers: innovators who buy the good independently of whom else have bought the good, and imitators who buy the good only if other customers have bought the good.

Section 2 investigates the dynamics of markets containing innovators only. Section 3 investigates markets with imitators. The main goal is to determine the temporal market evolution for various types of feedback from the market. One particularly important result is to determine the latency time, in this text, defined as the time it takes the market to reach 10% of all potential customers. This is a strategically important parameter since long latency time may result in premature closedown of services that eventually will become very lucrative.

Sections 2 and 3 study the evolution of the entire market without taking into consideration the effects of competition. Section 4 is about markets with several suppliers where the purpose is to study the evolution of the market toward stable equilibria for markets with and without churning. For markets with churning it is shown that for some churning functions the final state will consist of only one surviving supplier.

Section 5 presents models for interactive games. The effects of no feedback or positive feedback from the market for entering the game or quitting the game are studied. Altogether, six cases are discussed. The model consists of three states representing potential buyers (B), active players (P) and quitters (Q), respectively. The models consist of three coupled first order differential equations, one for each of the three states. One of the models is identical to the SIR (susceptible-infected-recovered) model of epidemiology (Section 5.2.3). The model in Section 5.2.2 allows a closed form analytic solution. The models in Sections 5.2.5 and 5.2.6 requires numerical solution of an integral, while the model in Sections 5.3.4 requires numerical integration of two coupled first order differential equations. The model in Section 5.2.7 requires numerical integration of a single first order differential equation.

Section 5.4 studies the effect of complementary games.

# 1. INTRODUCTION

## 1.1 Market dynamics

In a dynamic market model, temporal evolution of the market is studied. We are then, for example, concerned with how the number of subscribes to a service or players of a game evolves as a function of time. The dynamic behavior of the market can be described using analytic tools or simulation methods based on, for example, system dynamics (see, for example, [Ste]). System dynamics is a useful tool in cases where the market behavior is so complex that it is impractical or not even meaningful to use simple analytic tools. As with all simulation methods, system dynamics will not provide us with general solutions and thus general insight into the problem. The use of interacting agents is another simulation method that may be used to study the evolution of the market, see [Tes]. However, in this text, we shall only look at some simple market models where analytic solutions exist, and from which, important conclusions can be drawn.

We are only concerned with the dynamics of the market and not how factors such as price, availability, and ability to buy the good influence the desire to buy the good or subscribe to a service. The participants in the market are modeled as a uniform group where everyone has the same desire to buy the good or subscribe to the service. This is the same probabilistic method that is used in several fields in science and technology, for example, radioactive decay, epidemiology, teletraffic engineering, and population dynamics, just to mention a few.

The markets we are considering are markets where each customer buys at most one copy of each good. Typical examples of such markets are:

- subscription for telecommunications services (mobile phone, internet)
- newspapers, journals, and magazines,
- books, music, and films,
- games,
- subscription for energy,
- club memberships.

Other markets with a similar behavior are insurance and banking. In these cases, the customer may have similar contracts with more than one supplier, or have several contracts with the same supplier. The models are also applicable to the market for certain types of commodities such as refrigerators, furniture, and automobiles. These commodities have long working lives.

Let the number of potential users of a service (or game) be $u(t)$ at time $t$. The demand for the service during a short time interval $\Delta t$ is then the number of users $\Delta u$ buying the service during that interval; that is, the demand per unit time is defined as $\frac{\Delta u}{\Delta t}$ or, in the continuous limit, the demand per unit time becomes the time derivative $\frac{du}{dt}$, or $\dot{u}$ for short.[1]

The expressions for the number of customers having bought the good ($u$) and the time derivative of this expression ($\dot{u}$) will be computed for each model. The latter is then the number of customers buying the good per unit time. This is the same as the demand for the good.

---

[1] Following the normal convention in physics, we are using the terminology $\dot{u}$, $\ddot{u}$, $\dddot{u}$ ... (the dot derivative or Newton's derivative) for the time derivatives to distinguishing them from derivatives with respect to other variables. This enhances the readability of the formulas.

A market process such as buying or selling is a discrete process that may be described using discrete difference equations. However, the number of users is so large that we may approximate these difference equations as differential equations; that is, moving to the continuous limit (in the same way as is usually done in physics and mathematical biology). This provides us with a set of equations that are simpler to solve either analytically or numerically.

In a simple market where the customer may either be or not be a user at time $t$, the differential equation for the number of new users per unit time is then of the form

$$\dot{u} = f(u; t),$$

where $u = u(t)$ is the number of users at time $t$ and $f(u; t)$ is a function describing the market. The function may or may not contain time explicitly. Several examples of such functions depending or not depending explicitly on time will be studied below. This first order differential equation can then be solved (either analytically or numerically) for $u(t)$. This is then an expression for the number of users as a function of time $t$.

For more complex markets such as games, there may be several states representing, for example, potential buyers of the game, active players, and people who have quitted the game; or if there are several suppliers of the same good, there will be one differential equation for each supplier. Such cases can then be described by a set of linked first order differential equation, one equation for each state. For $k$ states, the set of equations is:

$$\dot{u}_1 = f_1(u_1, u_2, \cdots u_k; t),$$

$$\dot{u}_2 = f_2(u_1, u_2, \cdots u_k; t),$$

$$\vdots$$

$$\dot{u}_k = f_k(u_1, u_2, \cdots u_k; t).$$

In analogy to dynamic systems in physics, we may call $\dot{u}$ the velocity and $\ddot{u}$ the acceleration of the market. In the $k$-state system, the sets $\dot{\boldsymbol{u}} = (\dot{u}_1, \dot{u}_2, \dots \dot{u}_k)$ and $\ddot{\boldsymbol{u}} = (\ddot{u}_1, \ddot{u}_2 \dots \ddot{u}_k)$ may then be regarded as $k$-dimentional velocity vectors and acceleration vectors, respectively.

## 1.2 Market models

The basic market model was developed by Frank Bass during the 1960s [Bas]. The model is first of all describing the markets for commodities. The models contained in Sections 2 and 3 are variations on the Bass model. The normalized Bass model is

$$\dot{u}(t) = a(t)\big(1 - u(t)\big) + \gamma(t)(1 - u(t))u(t),$$

where the first term represents the buying behavior of people who are independent of whoever else has bought the good. The increase in the number of new customers at time $t$ is obviously proportional to the number of people who have not bought the good at that time. Bass called these customers *innovators*. The second term is proportional to both the number of potential buyers and the number of customers having bought the good. This represents a positive feedback from the market, or a network effect, or network externality. Bass called these customers *imitators*.

Normalization means that $u = U/N$ is the relative number of users, where $U$ is the absolute number of users and $N$ is the total population of potential users. The unnormalized Bass equation can then be written

$$\dot{U} = a(t)\big(N - U(t)\big) + \frac{\gamma(t)}{N}\big(N - U(t)\big)U(t).$$

Section 2 is about markets consisting of innovators only. These are markets where market feedback does not exist or is so weak that it may be neglected. The mobile phone market and the internet are examples of such markets. The basic model is

$$\dot{u}(t) = a(t)\big(1 - u(t)\big),$$

In Sections 2.4 to 2.6, the model is extended to include customers hesitating to buy the service and customers terminating the service.

Section 3 contains several models where there are different types of feedback from the market. The general model is

$$\dot{u}(t) = a(t)\big(1 - u(t)\big) + \gamma(t)\big(1 - u(t)\big)F\big(u(t)\big),$$

where $F\big(u(t)\big)$ is a general feedback term. Section 3 describes the full Bass model as well as several models without innovators, that is, models with market equations of the form

$$\dot{u}(t) = \gamma(t)\big(1 - u(t)\big)F\big(u(t)\big),$$

The effect of various forms of the feedback function $F\big(u(t)\big)$ is studied. This includes weak, medium and strong feedbacks, and the type of feedback that may be expected in markets dominated by trends.

Examples of telecommunications markets without innovators are SMS, email, facsimile and Facebook. These services are meaningless to a person if there is no one else with whom he or she may communicate. In other words, the desire to use the service depends on the number of people already using it.

Section 4 contains models where several suppliers compete for market shares. The Bass model (with variations) is used as the basic market model. In addition, the effect of churning is investigated, in particular to determine the long term market shares. One important question is to determine the final equilibrium state of the market. Churning may result in rather stable market shared between several suppliers such as the market for mobile communication, or in winner-take-all markets where one supplier will end up as a monopoly. Examples include markets where there are competing standards such as VHS and Betamax offering essentially the same service.

Games require more complex models. They are also, to some extent, based on the Bass model but are also related to epidemiological models (see for example [Hop], [Mur]). These models are considered in Section 5.

## 1.3 Equilibrium

One important definition is that of equilibrium points. The definition is particularly simple if the function $f(u; t)$ does not depend explicitly on time, that is, $f(u; t) = f(u)$. If the function depends explicitly on time, the equilibrium points may not be found by the method explained next. Examples 2 and 3 in Section 2.2 show two cases where the method fails.

The system is in equilibrium at the point $\tilde{u}$ if $f(\tilde{u}) = 0$. An equilibrium point is also a fixed point.[2] It is well-known from physics that we must have $\dot{u} = 0$ and $\ddot{u} = 0$ simultaneously at the equilibrium point; that is, both the velocity and the acceleration vanish at the equilibrium point. The latter condition is equivalent to the requirement that no net external force acts upon the system. Observe that if $f(u)$ is an everywhere smooth function of $u$,[3] then all the time derivatives of $u$ vanish at the equilibrium point $\tilde{u}$. In this case, the fixed points are equilibrium points.

For the $k$-dimentional case, the equilibrium points satisfy $\dot{\boldsymbol{u}} = (\dot{u}_1, \dot{u}_2, \ldots, \dot{u}_k) = (0,0,\ldots,0)$ and $\ddot{\boldsymbol{u}} = (\ddot{u}_1, \ddot{u}_2 \ldots, \ddot{u}_k) = (0,0,\ldots,0)$.

The equilibrium points may be stable or unstable. A stable equilibrium point is a point where the system will return to the equilibrium point after a small perturbation away from the point. Stable equilibrium points are called *attractors*. An unstable equilibrium point is a point where the system is at rest if no external forces act upon it. However, any perturbation (or small force) will cause the system to move away from the equilibrium point. Unstable equilibrium points are called *repellers*. There are also equilibrium points where some perturbations will cause the system to move back to equilibrium, while other perturbations will cause the system to move away from equilibrium. Such points are called *saddle points*. The saddle points are unstable. In the $k$-dimentional case, all points on a hypersurface of dimension less than $k$ may be equilibrium points, for example, if $k = 2$, all points on a line may be equilibrium points. We will encounter examples of such equilibrium points later.

## 1.4 Churning

If there are several suppliers of a service or good, churning may take place. Churning implies that a customer changes from one supplier to another at a certain time. Churning may then take place back and forth between different suppliers so long as the subscription lasts. The customer will still have just one subscription for the service (for example, mobile services) but obtaining the service from different suppliers at different times.

The churning function for supplier $i$ can be written in the form

$$C_i = \sum_{j \neq i} x_{ji}(u_j, u_i) - \sum_{j \neq i} x_{ij}(u_i, u_j),$$

where the first sum is the flow of customers from all suppliers $j$ to supplier $i$, and the second sum is the flow of customers from supplier $i$ to all other suppliers $j$. $C_i$ is then the net change in the number of customers of supplier $i$. $C_i$ may be negative or positive. Note that $\sum_i C_i = 0$ since churning does not alter the total number of customers but only the distribution of customers among the suppliers.

A reasonable assumption is that the rate by which a supplier is losing customers due to churning is proportional to the number of customers of that supplier since these are the potential churners. The churning function then takes the form

---

[2] A fixed point of a transformation $T$ (for example a motion) is a point $x$ where $T(x) = x$. In dynamical systems, an attractor or a repeller may be regarded as fixed points since no motion takes place at these points; we then regard motion as a transformation from one point in space to another. However, a fixed point may not be an equilibrium point. See also [Str].

[3] A function is everywhere smooth if the derivatives of any order of the function exist and are continuous at all points.

$$C_i = \sum_{j \neq i} a_{ji} u_j f_i(u_i, u_j) - u_i \sum_{j \neq i} a_{ij} f_j(u_j, u_i).$$

Here, $f_i(u_i, u_j)$ is the market feedback in favor of supplier $i$ (or popularity of supplier $i$) and the $a_{ij}$ are constants.

Churning will be studied further in Section 4.

## 2. MARKETS WITHOUT FEEDBACK

In Section 2 we shall look at models where there is only one supplier – or if there are several suppliers, the case where all the suppliers can be regarded as one uniform supplier. We are therefore not concerned with competition but with the way in which the market evolves as a whole. The markets for mobile services and internet may be modeled in this way: the evolution of the total market is independent of how many suppliers are offering the service and competing for market shares.

The present model is the simplest form of the Bass model where there are only innovators; that is, the desire to buy a product is independent of how many owns the product already.

In these models, the number of customers at time $t$ having bought the good is $S(t)$ and the demand is $D(t) = \dot{S}(t)$. For simplicity, we will use the normalized dependent variable $u(t) = S(t)/N$. The normalized initial condition is then $u_0 = S(0)/N$. The demand can be written as $D(t) = \dot{u}(t)N$.

### 2.1 All customers have the same constant adaptation rate $a$

Let $N$ be the initial total number of potential customers, and let $u(t)$ be the normalized number of consumers having subscription for a service, say, smartphone services, at time $t$. We assume that the adaptation rate, $a$, for the service is constant and the same for all potential customers. The rate at which the customers are applying for subscriptions is then

$$\dot{u}(t) = a(1 - u(t)).$$

This is so because the number of potentially new customers is $1 - u(t)$ after $u(t)$ customers already have subscribed to the service. These are then the customers that still may buy a subscription. The initial condition is $u(0) = u_0$. If $u_0 = 0$, no one has subscribed to the service initially.

We also see that $u = 1$ is an attractor (since $\dot{u} = \ddot{u} = 0$ for $u = 1$); the evolution of the market is such that eventually everyone becomes a subscriber.

The solution of this simple equation is

$$u(t) = 1 - (1 - u_0)e^{-at}.$$

The inverse of the adaptation rate is the average time between new subscriptions, $\tau$; that is, $\tau = 1/a$.

The demand $D(t) = N\dot{u}(t)$ is then $D(t) = aN(1 - u_0)e^{-at}$.

$u(t)$ is a monotonically increasing function of time, and $D(t)$ is a monotonically decreasing function of time. We also see that $u(\infty) = 1$, $D(0) = aN(1 - u_0)$, and $D(\infty) = 0$. Therefore, for small $t$, the number of customers having bought the good increases as $u(t) \cong at$. Moreover, since $u(\infty) = 1$, everyone subscribes to the service in the long term.

This is exactly the evolution of mobile services and internet we have observed in several countries.

If $u_0 = 0$, the time it takes until 50% of the market has been captured (that is, $u(t) = \frac{1}{2}$) is

$$T_{50} = \frac{\ln 2}{a},$$

or $a = \ln 2 / T_{50}$. The formula is useful since it is easy to select a suitable value for $T_{50}$, say 5 years, and from this value determine $a$ (for $T_{50} = 5$ years, $a = 0.14$ year$^{-1}$).

The latency time, defined here to be the time it takes until 10% of the market is captured, is

$$T_{10} = \frac{\ln(10/9)}{a} = 0.15 T_{50}.$$

This is a good measure for how fast the market increases initially, and is thus an important strategic parameter concerning whether services is likely to become lucrative in the long run.

If $T_{50} = 5$ years, then $T_{10} = 9$ months. This is a little faster than a linear evolution since for which, $T_{10}$ would be one year.

## 2.2 Time-dependent adaptation rate

If the adaption rate depends on time, that is, $a = a(t)$, then the equation for the number of subscribers becomes

$$\dot{u}(t) = a(t)(1 - u(t)).$$

We may expect that the attracting equilibrium point is $u = 1$ where the asymptotic solution ends up; however, as will be evident from Examples 2 and 3, this is not always the case.

For $u_0 = 0$, the solution is obviously[4]

$$u(t) = 1 - \exp\left[-\int_0^t a(x)dx\right].$$

The initial condition $u(0) = 0$ is fulfilled since $\int_0^0 a(x)dx = 0$ for a well-behaved function $a(u)$.

The demand is

$$D(t) = a(t)N \exp\left[-\int_0^t a(x)dx\right].$$

*Example* 1

For the linearly increasing adaptation rate $a(t) = a_0 + a_1 t$, we find:

---

[4] For the purpose of readability of formulas with complex expressions in the exponential, we use the notation $e^x = \exp(x)$.

$$u(t) = 1 - \exp\left[-a_0 t - \left(\frac{a_1}{2}\right)t^2\right].$$

We have $u(\infty) = 1$ so that this is an attracting equilibrium point.
The demand is

$$D(t) = (a_0 + a_1 t)N \exp\left[-a_0 t - \left(\frac{a_1}{2}\right)t^2\right].$$

*Example* 2

For the exponentially decreasing adaptation rate $a(t) = a_0 e^{-\beta t}$ we find:

$$u(t) = 1 - \exp\left[-\frac{a_0}{\beta}(1 - e^{-\beta t})\right].$$

The demand is

$$D(t) = a_0 N e^{-\beta t} \exp\left[-\frac{a_0}{\beta}(1 - e^{-\beta t})\right].$$

For $t = \infty$, we then have

$$u(\infty) = 1 - e^{-\frac{a_0}{\beta}} < 1;$$

that is, not everyone will become subscribers of the service. Hence, the point $u = 1$ is never reached even if it apparently is an equilibrium point. Therefore, we must be careful not drawing premature conclusions concerning stability when the differential equation depends explicitly on time. The solution approaches asymptotically another equilibrium point which cannot be determined from the differential equation directly. However, it is easily seen that both $\dot{u} = 0$ and $\ddot{u} = 0$ for $t = \infty$ so that the point $S(\infty) < N$ is, in fact, an equilibrium point.
On the other hand, $D(t) \to 0$ for $t \to \infty$ as it should.

*Example* 3

The adaptation rate is constant up to time $T$; thereafter, the adaptation rate is zero; that is,

$$a(t) = \begin{cases} a, & t \leq T \\ 0, & t > T \end{cases}$$

This gives

$$u(t) = \begin{cases} 1 - e^{-at}, & t \leq T \\ 1 - e^{-aT}, & t > T \end{cases}$$

This example is also a case where the asymptotic solution does not correspond to the expected equilibrium point. For the demand, we find

$$D(t) = \begin{cases} aNe^{-at}, & t \leq T \\ 0, & t > T \end{cases}.$$

## 2.3 Segmented markets

Figure 2.1 shows a model of a market segmented into $k$ segments $N_1, N_2, \cdots, N_k$ with constant adaptation rates $a_1, a_2, \cdots, a_k$, respectively. The total number of potential subscribers is then $N = \sum_1^k N_i$. We normalize as before, setting $n_i = N_i/N$ and $u_i = S_i/N$.

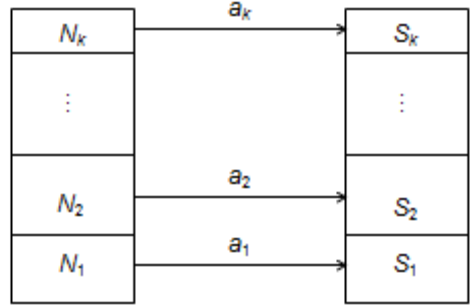

Figure 2.1 Segmented market

All segments are independent. The demand from customers in segment $N_i$ is then

$$\dot{u}_i(t) = a_i(n_i - u_i(t))$$

with solution

$$u_i(t) = n_i(1 - e^{-a_i t}).$$

The behavior of the total market is therefore ($u_0 = 0$)

$$u(t) = \sum_1^k u_i(t) = 1 - \sum_1^k n_i e^{-a_i t},$$

and the demand is

$$D(t) = \sum_1^k a_i N_i e^{-a_i t}.$$

For the more general case where the $a_i$ depends on time, we find

$$u(t) = 1 - \sum_1^k n_i \exp[-\int_0^t a_i(x)dx],$$

$$D(t) = \sum_1^k a_i(t) N_i \exp[-\int_0^t a_i(x)dx].$$

## 2.4 Adaptation with hesitation 1

The model shown in Figure 2.2 contains three categories of customers: potential adapters $P$, hesitant or delayed adapters $H$, and users $U$. In normalized variables: $p = P/N$, $h = H/N$ and $u = S/N$. This is then a three-stage model. The potential adapters may either subscribe to the service with intensity $a$, or enter the hesitating state with intensity $b$. Those in the hesitating state will then subscribe to the service with intensity $c$, where $c$ should be smaller than $a$. Average hesitation time is then $\tau_h = \frac{1}{b} + \frac{1}{c}$.

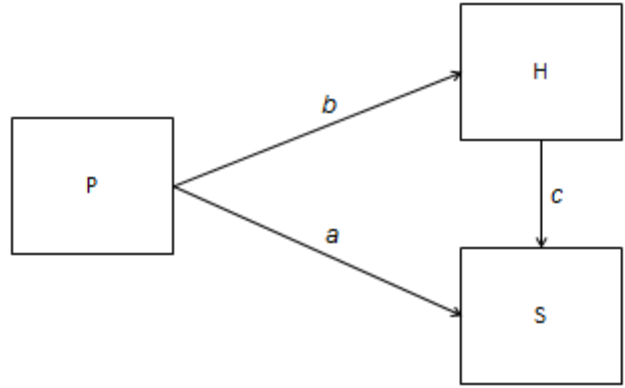

Figure 2.2 Model with hesitant customers

The set of differential equations describing the temporal behavior of $p$, $h$, and $s$ is:

$$\dot{p} = -ap - bp,$$

$$\dot{u} = ap + ch,$$

$$\dot{h} = bp - ch.$$

We see immediately that $p + u + h = 1$. Moreover, we may assume that there are no initial customers, that is, $p(0) = 1, u(0) = 0, h(0) = 0$. The solution for $p$ is readily found:

$$p = e^{-(a+b)t}.$$

From $p + u + h = 1$ we also see that $h = 1 - p - u = 1 - e^{-(a+b)t} - u$, giving

$$\dot{u} = ae^{-(a+b)t} + c\left[1 - e^{-(a+b)t}\right] - cu.$$

The solution of this inhomogeneous linear equation is readily found using the method of integrating factor [Inc], [Kor].

$$u(t) = 1 - \frac{1}{a+b-c}\left[be^{-ct} + (a-c)e^{-(a+b)t}\right].$$

The demand is then

$$D(t) = \frac{N}{a+b-c}\left[bce^{-ct} + (a-c)(a+b)e^{-(a+b)t}\right].$$

## 2.5 Adaptation by hesitation 2

In this model we assume that a customer entering the hesitation state may go back to the potential subscriber state with intensity $c$ as shown in Figure 2.3. We are using the same normalized variables as in Section 2.4.

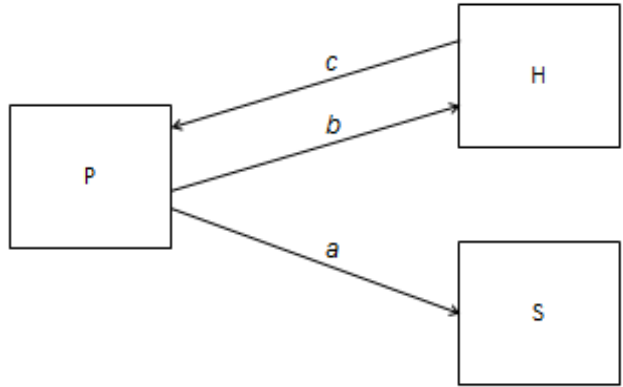

Figure 2.3 Second model with hesitating customers

The differential equations are now:

$$\dot{p} = -ap - bp + ch,$$

$$\dot{u} = ap,$$

$$\dot{h} = bp - ch.$$

First, we observe that the equation for $u$ is

$$u(t) = a \int_0^t p(x)dx.$$

and that $p$ and $h$ are determined by the first and the last equation.

Eliminating $p$ and $h$ from these equations will lead to a new second order differential equation (that can be readily solved since it is linear with constant coefficients). However, we will solve the equations using a more elegant method only using first order derivatives. The different steps in the procedure are shown without proof, see [Goe] or [Kor]. The equations for $p$ and $h$ can be written in matrix form

$$\begin{pmatrix} \dot{p} \\ \dot{h} \end{pmatrix} = \begin{pmatrix} -a-b & c \\ b & -c \end{pmatrix} \begin{pmatrix} p \\ h \end{pmatrix} = M \begin{pmatrix} p \\ h \end{pmatrix},$$

where $M = \begin{pmatrix} -a-b & c \\ b & -c \end{pmatrix}$. Formally, treating the vector $\begin{pmatrix} p \\ h \end{pmatrix} = V$ as a single object, we may write the equation in the form $\dot{V} = MV$ or $\frac{dV}{V} = Mdt$, and seek the solution of this equation in the form

$$V = \begin{pmatrix} p \\ h \end{pmatrix} = e^{Mt}V_0 = e^{Mt}\begin{pmatrix} p(0) \\ h(0) \end{pmatrix},$$

where

$$e^{Mt} \stackrel{\text{def}}{=} 1 + \frac{Mt}{1!} + \frac{(Mt)^2}{2!} + \frac{(Mt)^3}{3!} + \cdots$$

The exponential of a matrix is itself a matrix since the right-hand side of the defining equation consists only of the elementary matrix operations addition and multiplication. The matrix $e^{Mt}$ can be evaluated by diagonalization as follows.

First we find the *eigenvalues* of the matrix. These are the roots $\lambda_1$ and $\lambda_2$ of the determinantal equation

$$\det\begin{pmatrix} -a - b - \lambda & c \\ b & -c - \lambda \end{pmatrix} = 0.$$

This gives the quadratic equation $(a + b + \lambda)(c + \lambda) - bc = 0$ with roots $2\lambda_{1,2} = -(a + b + c) \pm \sqrt{(a + b + c)^2 - 4ac}$.

We easily see that the term under the square root sign is positive for all positive $a$, $b$, and $c$, and that the square root is obviously less than $a + b + c$.[5] Hence, the *eigenvalues* are real and negative. Moreover, $\lambda_2 < \lambda_1 < 0$.

The motivation for this procedure is that any diagonalizable matrix $M$ can be written in the form

$$M = A^{-1} \begin{pmatrix} \lambda_1 & 0 \\ 0 & \lambda_2 \end{pmatrix} A = A^{-1} \Lambda A$$

where $A$ is a $2 \times 2$ matrix.[6] Inserting this in the defining equation for $e^{Mt}$ gives

$$e^{Mt} = \sum_0^\infty \frac{(Mt)^i}{i!} = A^{-1} \left( \sum_0^\infty \frac{(\Lambda t)^i}{i!} \right) A = A^{-1} \begin{pmatrix} \sum_0^\infty (\lambda_1 t)^i / i! & 0 \\ 0 & \sum_0^\infty (\lambda_2 t)^i / i! \end{pmatrix} A,$$

that is,

$$e^{Mt} = A^{-1} \begin{pmatrix} e^{\lambda_1 t} & 0 \\ 0 & e^{\lambda_2 t} \end{pmatrix} A$$

To find the matrix $A$, we have first to determine the *eigenvectors* $\begin{pmatrix} v_{11} \\ v_{21} \end{pmatrix}$ and $\begin{pmatrix} v_{12} \\ v_{22} \end{pmatrix}$ corresponding to the two *eigenvalues*:

$$\begin{pmatrix} -a - b & c \\ b & -c \end{pmatrix} \begin{pmatrix} v_{11} \\ v_{21} \end{pmatrix} = \lambda_1 \begin{pmatrix} v_{11} \\ v_{21} \end{pmatrix},$$

$$\begin{pmatrix} -a - b & c \\ b & -c \end{pmatrix} \begin{pmatrix} v_{12} \\ v_{22} \end{pmatrix} = \lambda_2 \begin{pmatrix} v_{12} \\ v_{22} \end{pmatrix}.$$

The matrix $A$ above is then given by $A = \begin{pmatrix} v_{11} & v_{12} \\ v_{21} & v_{22} \end{pmatrix}$.

Each matrix equation gives rise to two linear equations for the *eigenvectors*; however, the two equations are not independent – simple algebra shows, in fact, that the equations are identical. Therefore, we arbitrarily choose one independent equation for each *eigenvector*, for example,

$$bv_{11} - cv_{21} = \lambda_1 v_{21},$$

$$bv_{12} - cv_{22} = \lambda_2 v_{22}.$$

The *eigenvectors* cannot be determined uniquely; however, one suitable set of *eigenvectors* is $v_{11} = c + \lambda_1$, $v_{21} = b$, $v_{12} = c + \lambda_2$, and $v_{22} = b$.

The solutions for $p$ and $h$ are then

---

[5] We have $0 \le (a + b - c)^2 = (a + b + c - 2c)^2 = (a + b + c)^2 - 4ac - 4bc \le (a + b + c)^2 - 4ac$.

[6] Note that the same procedure is valid for any diagonalizable $n \times n$ matrix for any $n$.

$$\begin{pmatrix} p \\ h \end{pmatrix} = \alpha \begin{pmatrix} v_{11} \\ v_{21} \end{pmatrix} e^{\lambda_1 t} + \beta \begin{pmatrix} v_{12} \\ v_{22} \end{pmatrix} e^{\lambda_2 t},$$

that is,

$$p = \alpha(c + \lambda_1)e^{\lambda_1 t} + \beta(c + \lambda_2)e^{\lambda_2 t},$$

$$h = \alpha b e^{\lambda_1 t} + \beta b e^{\lambda_2 t}.$$

The constants of integration $\alpha$ and $\beta$ are determined from the initial conditions:

$$p(0) = 1 \Rightarrow 1 = \alpha(c + \lambda_1) + \beta(c + \lambda_2),$$

$$h(0) = 0 \Rightarrow 0 = \alpha b + \beta b.$$

The constants of integration are then

$$\alpha = -\beta = \frac{1}{\lambda_1 - \lambda_2}.$$

This gives

$$p = \frac{1}{\lambda_1 - \lambda_2}[(c + \lambda_1)e^{\lambda_1 t} - (c + \lambda_2)e^{\lambda_2 t}]$$

and

$$h = \frac{b}{\lambda_1 - \lambda_2}[e^{\lambda_1 t} - e^{\lambda_2 t}].$$

Finally, the equation for the number of subscribers is

$$u = a\int_0^t p(x)dx = \frac{a}{(\lambda_1 - \lambda_2)\lambda_1\lambda_2}[\lambda_2(c + \lambda_1)e^{\lambda_1 t} - \lambda_1(c + \lambda_2)e^{\lambda_2 t}] + \frac{ac}{\lambda_1\lambda_2}$$

Setting $r = \sqrt{(a + b + c)^2 - 4ac}$, then $2\lambda_1 = -(a + b + c) + r$, and $2\lambda_2 = -(a + b + c) - r$; that is, $\lambda_1 - \lambda_2 = r$ and $\lambda_1\lambda_2 = ac$. The formula for $u$ can then be written in the form

$$u(t) = 1 - \frac{1}{r}e^{\lambda_1 t}[(a + \lambda_1)e^{-rt} - a - \lambda_2].$$

The demand is

$$D(t) = \frac{Nc}{r}e^{\lambda_1 t}[a + \lambda_1 - (a + \lambda_2)e^{-rt}].$$

## 2.6 Model with birth and death rates but without hesitation

The model in Figure 2.4 shows the case where the number of potential subscribers, $P$, increases at rate $d$ and decreases at rate $f$, while the number of subscribers decreases at rate $g$; $d$, and $f$ and $g$ may be called birth and death rates, respectively. Moreover, we assume that $a + f > d + g$. The differential equations are:

$$\dot{p} = dp - ap - fp,$$

$$\dot{u} = ap - gu.$$

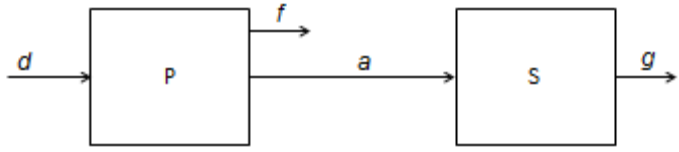

Figure 2.4 Model with birth and death rates

Here, we again determine $p$ from the first equation and, from that solution, deduce the equation for $u$:

$$\dot{u} = ae^{-(a+f-d)t} - gu.$$

This equation is a non-homogeneous equation which is solved using the method of integrating factor. The solution is (for $u_0 = 0$):

$$u(t) = \frac{a}{a+f-d-g} e^{-gt}\left[1 - e^{-(a+f-d-g)t}\right].$$

$$D(t) = aNp = aNe^{-(a+f-d-g)t}.$$

## 3. MARKETS WITH FEEDBACK

Feedback from the market implies that the attractiveness of a product depends on the number of users of the product. The feedback is also referred to as a network effect or network externality [Sha].

We start with studying the Bass model. The effect of the strength of the feedback is also investigated in the case when all users are imitators. In Section 3.2, we consider network effects where the attractiveness of the product decreases as a function of customers, illustrating the case where the market stagnates after a fierce initial increase.

The general market equation is:

$$\dot{u} = a(1-u)F(u),$$

where $F(u)$ is the feedback term: $F(u) = 1$ means that there is no feedback. Moreover, all coupling parameters are constants and the feedback term does not contain time explicitly. Then the differential equation is separable with general solution

$$\int \frac{du}{(1-u)F(u)} = at + c,$$

where $c$ is a constant of integration determined by the initial value for $u = u(0) = u_0$.

The solutions for the different models are compared by requiring that the time it takes to capture 50% of the market, $T_{50}$, is the same in all models. This parameter is a simple strategic decision variable where, for example, a service is implemented provided that it takes no more than five years to capture 50% of the market. Otherwise the service is terminated.

## 3.1 Bass model

### 3.1.1 Both innovators and imitators

The model is the diffusion model of Frank Bass (see Section 1.2). The differential equation can then be written $\dot{u} = a(1-u) + \gamma u(1-u)$. Brass calls $a$ the coefficient of *innovation* and $\gamma$ the coefficient of *imitation*. The feedback term is then $F(u) = 1 + (\gamma/a)u$.

The first term is the adaptation rate of innovators; the second term is the adaption rate of imitators. Note that this equation has only one equilibrium point ($u = 1$) for positive $u$; this point is an attractor.

Separating the variables gives:

$$\frac{du}{(a+\gamma u)(1-u)} = dt.$$

This is the same as

$$\frac{\gamma du}{a+\gamma u} + \frac{du}{1-u} = (a+\gamma)dt.$$

The solution is

$$u = \frac{a + \gamma u_0 - a(1-u_0)e^{-(a+\gamma)t}}{a + \gamma u_0 + \gamma(1-u_0)e^{-(a+\gamma)t}},$$

where $u(0) = u_0$. We also see that the market starts growing also for $u_0 = 0$. The equation is then:

$$u = \frac{1 - e^{-(a+\gamma)t}}{1 + (\gamma/a)\, e^{-(a+\gamma)t}}.$$

Solved for $t$, this gives

$$t = \frac{1}{a+\gamma} \ln \frac{1 + (\gamma/a)\, u}{1-u}.$$

This gives for $T_{50}$ and $T_{10}$:

$$T_{50} = \frac{1}{a+\gamma} \ln\left(2 + \frac{\gamma}{a}\right)$$

$$T_{10} = \frac{1}{a+\gamma} \ln\left(1.11 + 0.11\frac{\gamma}{a}\right) = T_{50} \frac{\ln\left(1.11 + 0.11\frac{\gamma}{a}\right)}{\ln\left(2 + \frac{\gamma}{a}\right)}$$

The sum $a + \gamma$ is given by $a + \gamma = \frac{1}{T_{50}} \ln\left(2 + \frac{\gamma}{a}\right)$. In this case, we may choose a value for the ratio $\gamma/a$ and compute $a + \gamma$ from the above formula.

Sales per unit time is

$$D(t) = \frac{N(a+\gamma)^2(1-u_0)(a+\gamma u_0)e^{-(a+\gamma)t}}{[a + \gamma u_0 + \gamma(1-u_0)e^{-(a+\gamma)t}]^2}.$$

The inflexion point (if it exists for positive $u$) corresponds to the case where the sales per unit time is maximum. This gives $\ddot{u} = 0 = (-a + \gamma)\dot{u} - 2\gamma u\dot{u}$. Solved for $u$ we find $u_{infl} = (\gamma - a)/2\gamma$.

The inflexion point exists for positive $u$ if $\gamma > a$. If we set $u_0 = 0$, the inflexion point (the maximum sales rate) is reached after time $t_{infl} = (\ln \gamma - \ln a)/(\gamma + a)$.

### 3.1.2 Only imitators: Linear positive feedback

This case corresponds to the Bass model with imitators only. This model applies to services such as SMS, telefax and Facebook where there is no reason to subscribe to the service unless there is at least one other subscriber to communicate with.

The differential equation for the market evolution is then

$$\dot{u} = \gamma u(1 - u).$$

The feedback term is now $F(u) = u$.

The coupling factor $\gamma u(t)$ represents linear[7] positive feedback from the market; that is, if the number of customers increases, more people are stimulated to buy the product.[8] This type of market feedback is sometimes also called a *network externality*. The equation is called the *logistic differential equation*.[9]

The differential equation that can be written

$$\frac{du}{u(1-u)} = \frac{du}{u} + \frac{du}{1-u} = \gamma dt.$$

The solution is

$$u(t) = \frac{u_0}{u_0 + (1 - u_0)e^{-\gamma t}},$$

where $u_0 = u(0)$ is the initial value of $u(t)$.

Solved for $t$, we find

$$t = \frac{1}{\gamma} \ln \frac{u(1 - u_0)}{u_0(1 - u)}.$$

The parameter $\gamma$ is then estimated from this formula for $u = \frac{1}{2}$ and $t = T_{50}$. This gives

---

[7] Since the feedback function is of the first order, or linear, in $u$.

[8] In a system with *positive feedback*, a small perturbation in the output from the system will result in a bigger perturbation in the output. The result is a runaway system where the output from the system increases toward saturation, becomes empty, or oscillates in a regular or irregular manner. Oscillations may occur if the feedback signal is delayed, for example, in markets where the amount produced must be chosen before the prices are known, or in education where the expected future demand for professionals in a certain field is based on the demand when the training starts. In a system with *negative feedback*, the feedback will counteract any perturbation in the output such that the perturbation is reduced, or, in other words, the negative feedback stabilizes the output from the system.

[9] There is also a logistic difference equation, which for particular choices of parameter $a$, gives rise to chaotic solutions with bifurcations; see for example [Str] and [Sch]. This equation is not important for the markets studied in this text.

$$\gamma = \frac{1}{T_{50}} \ln \frac{1-u_0}{u_0}.$$

The inflexion point $T_{infl}^{lin}$ of the function $u$ is the point where $\ddot{u} = 0$. Since $\dot{u} = \gamma u(1-u)$ it follows that $\ddot{u} = (1-2u)\dot{u}$; that is, the coordinates for the inflexion point are $u_{infl} = \frac{1}{2}$ and $T_{infl} = T_{50}$. Moreover, the gradient at the inflexion point is:

$$u_{infl}^{lin} = \frac{1}{4T_{50}} \ln \frac{1-u_0}{u_0}.$$

For $u_0 = 0.01$, this gives

$$u_{infl}^{lin} = \frac{1.15}{T_{50}}.$$

The latency time is (again taken as the time it takes to capture 10% of the market)

$$T_{10} = T_{50} \frac{\ln[0.1(1-u_0)] - \ln[0.9u_0]}{\ln(1-u_0) - \ln u_0}.$$

Table 1 shows $\frac{T_{10}}{T_{50}}$ for some values of $u_0$ and $T_{50} = 5$ years.

Table 3.1 Latency time for some values of $u_0$

| $u_0$ | $T_{10}/T_{50}$ | $T_{10}$ for $T_{50} = 5$ years |
|---|---|---|
| 0.001 | 0.67 | 3 years and 4 months |
| 0.005 | 0,58 | 2 years and 11 months |
| 0.01 | 0.52 | 2 years and 7 months |
| 0.02 | 0.44 | 2 years and 2 months |
| 0.04 | 0.31 | 1 year and 6 months |

Observe that if $u_0 = 0$ then $u(t) = 0$ for all $t$; that is, the customers will buy the service only if there already are customers who have bought the service. This is one of the strategic difficulties in markets with positive feedback and no innovators: the supplier must in one way or another establish an initial customer base before the service is launched, for example, offering the service for free to some trial customers. Examples of information services with positive feedback are SMS, Facebook, LinkedIn, telephone service, data communication, telefax, and interactive games. Note that there is no significant positive feedback for mobile telephony since the customers of mobile services can communicate with customers of the fixed network; there is no significant positive feedback for MMS since SMS is a subset of this service with an established customer base.

Table 3.1 illustrates a difficult strategic problem, namely the time it takes for the market to increase to an acceptable level (the latency time). If it is expected that the product will be bought by 50 percent of the potential customers after 5 years, the table illustrates that the supplier may face a severe problem since the market share is only 10 percent after almost three years for an initial customer base is 0.5%. If the initial market share is 2%, it takes more than two years to reach 10% market share. This means that the supplier wrongly may conclude that the market share of 50% after 5 years is overoptimistic and, therefore, withdraw the product from the market.

For the logistic equation, the equilibrium points are the solutions of the equation $\dot{u} = \gamma u(1 - u) = 0$; that is, $u = 0$ or $u = 1$. Note that both the velocity and acceleration vanishes for these values of $u$, so that these points are also equilibrium points.

The point $u(0) = 0$ is an unstable equilibrium point (or repeller) since any perturbation away from the point will cause $u(t)$ to grow. The point $u(\infty) = 1$ is a stable equilibrium point (or attractor) since any perturbation away from 1 will cause the system to return to 1.

The solution for $u_0 > 0$ is a sigmoid or S-curve as illustrated in Figure 3.1.

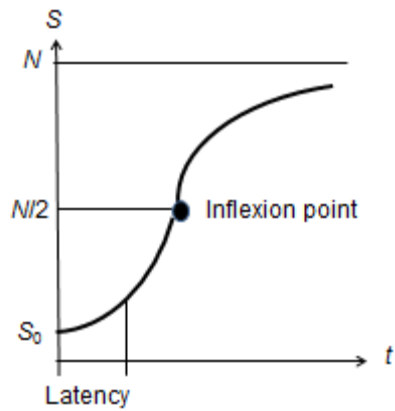

Figure 3.1 Sigmoid ($S = Nu$ as a function of time)

The demand is

$$D(t) = \frac{N\gamma u_0 e^{-\gamma t}}{[u_0 + (1 - u_0)e^{-\gamma t}]^2}.$$

We see that the initial demand is

$$D(0) = N\gamma u_0 = \gamma S(0).$$

### 3.1.3 Only imitators: Weak feedback

In this section, the effect of a weaker feedback $\sqrt{u}$ is explored. The differential equation is now $\dot{u} = \gamma_{1/2}\sqrt{u}\,(1 - u)$ with feedback term $F(u) = \sqrt{u}$.

The differential equation is

$$\frac{du}{\sqrt{u}(1 - u)} = \gamma_{1/2}dt.$$

Introducing a new dependent variable $v^2 = u$, we get

$$\frac{2dv}{1 - v^2} = \frac{dv}{1 + v} + \frac{dv}{1 - v} = \gamma_{1/2}dt.$$

Integration gives

$$\frac{1 + v}{1 - v} = ce^{\gamma_{1/2}t},$$

where the constant of integration, satisfying the initial condition $u(0) = u_0$, is

$$c = \frac{1 + \sqrt{u_0}}{1 - \sqrt{u_0}}$$

Solving for $u$ gives, after some simple calculations,

$$u = \left[\frac{1 + \sqrt{u_0} - (1 - \sqrt{u_0})e^{-\gamma_{1/2}t}}{1 + \sqrt{u_0} + (1 - \sqrt{u_0})e^{-\gamma_{1/2}t}}\right]^2.$$

As we shall see, $u(0) = u_0 = 0$ is not an equilibrium point. The reason for such behavior is that $\sqrt{u}$ is not regular at the point $u = 0$ (the point is a branch point separating the positive and the negative branch of the square root). The growth curve starting from $u(0) = u_0 = 0$ is then

$$u = \left[\frac{1 - e^{-\gamma_{1/2}t}}{1 + e^{-\gamma_{1/2}t}}\right]^2.$$

The second derivative at $u(0) = 0$ is easily found by derivation of the original equation:

$$\ddot{u} = \gamma_{1/2}\left[\frac{\dot{u}}{2\sqrt{u}}(1 - u) - \dot{u}\sqrt{u}\right].$$

Inserting $\dot{u}/\sqrt{u} = \gamma_{1/2}(1 - u)$ and setting $u = 0$, we get $\ddot{u}(0) = \frac{1}{2}\gamma_{1/2}^2 > 0$. Since the acceleration at $u(0) = 0$ is nonzero, $u(0) = 0$ is not an equilibrium point.

The inflexion points are given by $\ddot{u} = 0$, that is,

$$\ddot{u} = \gamma_{1/2}\left[\frac{\dot{u}}{2\sqrt{u}}(1 - u) - \dot{u}\sqrt{u}\right] = 0,$$

with the single solution ($\dot{u} \neq 0$) $u_{infl}^{sqr} = \frac{1}{3}$.

The inflexion point is located at a smaller $u$ than for linear positive feedback. The gradient at the inflexion point is (using the formula for $T_{50}$ computed below):

$$\dot{u}_{infl}^{sqr} = \frac{2}{3\sqrt{3}}\gamma_{1/2} = \frac{0.68}{T_{50}}.$$

Solving the equation for $t$ gives

$$t = \frac{1}{\gamma_{1/2}}\ln\frac{(1 - \sqrt{u_0})(1 + \sqrt{u})}{(1 + \sqrt{u_0})(1 - \sqrt{u})}.$$

The 50% point is found by setting $u = \frac{1}{2}$. This gives for $u(0) = 0$

$$T_{50} = \frac{1}{\gamma_{1/2}}\ln\frac{\sqrt{2} + 1}{\sqrt{2} - 1} = \frac{1.76}{\gamma_{1/2}}.$$

The growth parameter $\gamma_{1/2}$ is given by

$$\gamma_{1/2} = \frac{1.76}{T_{50}}.$$

The 10% latency time is found by setting $u = 0.1$. This gives

$$T_{10} = \frac{1}{\gamma_{1/2}} \ln \frac{\sqrt{10}+1}{\sqrt{10}-1} = \frac{0.655}{\gamma_{1/2}},$$

or if $u_0 = 0$, $T_{10} = 0{,}37 T_{50}$; that is, 1 year and 10 months if $T_{50} = 5$ years. For $u_0 = 0.01$ we find $T_{10} = 0.29 T_{50}$, or 1 year and 5 months.

The demand is

$$D(t) = \frac{4N(1-u_0)\gamma_{1/2} e^{-\gamma_{1/2} t}\left[1+\sqrt{u_0}-(1-\sqrt{u_0})e^{-\gamma_{1/2} t}\right]}{\left[1+\sqrt{u_0}+(1-\sqrt{u_0})e^{-\gamma_{1/2} t}\right]^3},$$

$$D(0) = N\gamma_{1/2}(1-u_0)\sqrt{u_0}.$$

### 3.1.4 Only imitators: Strong feedback

If the feedback is proportional to the square of the number of subscribers, the market equation is $\dot{u} = \gamma_2 u^2(1-u)$. The feedback term is $F(u) = u^2$. This equation is used to study the effect of a strong positive feedback from the market. The differential equation is:[10]

$$\frac{du}{u^2(1-u)} = \frac{du}{u} + \frac{du}{u^2} + \frac{du}{1-u} = \gamma_2 dt.$$

Integrating term by term gives:

$$\ln u - \frac{1}{u} - \ln(1-u) = \ln \frac{u e^{-\frac{1}{u}}}{1-u} = \gamma_2 t + \hat{c}.$$

For an initial value $u(0) = u_0$, the solution is

$$\frac{u}{1-u} e^{-\frac{1}{u}} = \frac{u_0}{1-u_0} \exp\left(\gamma_2 t - \frac{1}{u_0}\right).$$

This is a transcendental equation that cannot be solved explicitly for $u$; however, the solution is easily plotted using $t$ as dependent variable:

$$t = \frac{1}{\gamma_2} \ln\left[\frac{u(u-u_0)}{(1-u)u_0} \exp\left(\frac{1}{u_0}-\frac{1}{u}\right)\right] = \frac{1}{\gamma_2}\left[\ln\frac{u(u-u_0)}{(1-u)u_0} + \frac{1}{u_0} - \frac{1}{u}\right]$$

We see that $u = 0$ is a repelling equilibrium point and $u = 1$ is an attracting equilibrium point. This also implies, just as for linear feedback, $u_0 > 0$ for a non-zero solution to emerge.

The time $T_{50}$ is given by the formula

$$\gamma_2 T_{50} = \ln \frac{1-2u_0}{2u_0} + \frac{1}{u_0} - 2,$$

---

[10] Note that the inverse of a polynomial factored into real factors can be written as a sum over the inverse factors, for example, $\frac{1}{(x-a)(x-b)^2(x-c)(x^2+dx+e)}$ can be written as $\frac{\alpha}{x-a} + \frac{\beta}{x-b} + \frac{\gamma}{(x-b)^2} + \frac{\delta}{x-c} + \frac{\varepsilon x+\zeta}{x^2+dx+e}$ for unique values of $\alpha, \beta$, etc, only dependent on the constants $a, b$, etc. Note how multiple factors such as $(x-b)^2$ and nonlinear terms are handled.

and, similarly for $T_{10}$:

$$\gamma_2 T_{10} = \ln\frac{0.1 - u_0}{u_0} + \frac{1}{u_0} - 12.2.$$

For $u_0 = 0.01$, we get $T_{10} = 0.88 T_{50}$. If $T_{50} = 5$ years, then $T_{10} = 4$ years and 4 months. This implies that 10% of the market is taken after 4 years and 4 months, while the next 40% of the market is taken during the next eight months. This illustrates clearly the strategic dilemma related to markets with positive feedback.

The inflexion point is given by the equation $\ddot{u} = 0 = \gamma \dot{u} u(2 - 3u)$. This gives $u_{infl}^{quadr} = \frac{2}{3}$.

In this case, the inflexion point is located at a larger value of $u$ compared to linear positive feedback. The gradient at $\dot{u}_{infl}^{quadr}$ is $\dot{u}_{infl}^{quadr} = \frac{4}{27}\gamma_2 = \frac{1}{T_{50}}(\ln\frac{1-2u_0}{2u_0} + \frac{1}{u_0} - 2)$.

Note that it is not possible to derive an analytic expression for the demand as a function of time since this would require that $u$ is expressible as a function of time.

### Note 1

Note the following generalization. The differential equation for feedback terms proportional to $u^n$ where $0 \leq n < \infty$, is $\dot{u} = \gamma_n u^n(1 - u)$. Taking the second derivative, we find $\ddot{u} = \gamma_n \dot{u} u^{n-1}[n - (n+1)u]$. Hence, the inflexion point is located at $u_{infl} = \frac{n}{n+1}$.

Table 3.2 shows the location of the inflexion point for some values of $n$ and for fixed 50%-point.

Table 3.2 Location of inflexion point

| $n$ | Inflexion point | Relative to $T_{50}$ |
|---|---|---|
| 0.1 | 1/11 | <$T_{50}$ |
| 0.5 | 1/3 | <$T_{50}$ |
| 1 | 1/2 | =$T_{50}$ |
| 2 | 2/3 | >$T_{50}$ |
| 10 | 10/11 | >$T_{50}$ |

Form the table, the differential equation and the definition of an inflexion point, we may conclude
- that the smaller $n$ is, the shorter is the latency time; the latency time approaches that of the simple market (Section 2.1) for $n \to 0$;
- that the bigger $n$ is, the longer is the latency time, and the curve is approaching a step function at the 50%-point as $n$ increases toward infinity ($T_{10} \to T_{50}$ for $n \to \infty$).

### Note 2

It is easy to find analytic solutions of the differential equation $\dot{u} = \gamma_n u^n(1 - u)$ for several values of $n$ other than those considered here ($\frac{1}{2}$, 1 and 2), for example, $n = \frac{1}{8}, \frac{1}{4}, \frac{1}{3}$, and all integer $n$.

## 3.2 Decreasing network effects – modeling a trend

### 3.2.1 Attractiveness proportional to the number of non-subscribers

The idea here is that the product is more popular the fewer who owns it; that is, the feedback is proportional to $1 - u$. The differential equation is $\dot{u} = \gamma_0(1 - u)^2$ with feedback term $F(u) = 1 - u$, or

$$\frac{du}{(1-u)^2} = \gamma_0 dt$$

The solution satisfying $u = 0$ for $t = 0$ is

$$u = \frac{\gamma_0 t}{\gamma_0 t + 1}.$$

We see also that $u = 1$ for $t = \infty$. Moreover,

$$T_{50} = \frac{1}{\gamma_0} \text{ and } T_{10} = \frac{T_{50}}{9}.$$

If $T_{50} = 5$ years, then $T_{10}$ is 6 months and 18 days. This is just a little shorter than in a market without feedback. In Section 2.1, we found that $T_{10}$ is 9 months. Let us then look at feedbacks that reduces the latency time further.

### 3.2.2 Attractiveness inversely proportional to the number of subscribers

Assume that the demand equation has the form

$$\dot{u} = \gamma_{-1} \frac{1-u}{u}$$

with feedback term $F(u) = 1/u$.

This implies that the attractiveness of the product is infinite when the product is marketed, and then gets smaller as more and more customers buy the good. A trend may be modeled this way. One characteristic of a trend is that there are some users adapting the product when it is new and few owns it, but then loses the interest in it when the product becomes common.

The differential equation can be written

$$\frac{udu}{1-u} = -du + \frac{du}{1-u} = \gamma_{-1} dt.$$

This gives a transcendental equation for $u$:

$$e^u(1-u) = e^{-\gamma_{-1}t},$$

or

$$t = -\frac{1}{\gamma_{-1}}[u + \ln(1-u)].$$

We observe that if $t = 0$, then $u = 0$, and if $t = \infty$, then $u = 1$.[11] We also see from the original equation that the tangent to the curve at $t = 0$ is vertical, so that the market rises extremely fast initially for later to flatten out. The constant $\gamma_{-1}$ is determined from the time it takes to reach 50% of the market. We find easily:

$$\gamma_{-1} = \frac{0.19}{T_{50}}.$$

---

[11] If $t = 0$, then $e^u(1-u) = 1$ with solution $u = 0$. If $t \to \infty$, $e^u(1-u) = 0$ with solution $u = 1$.

The 10% latency time is easily found: $T_{10} = 0.028 T_{50}$. The latency time is very small. If $T_{50} = 5$ years, then $T_{10} = 1$ month and 20 days. Similarly, we find that it takes a little more than 7 month to reach 20% of the market. This illustrates that trends also embody strategic dilemmas. The initial increase is so violent that the future increase may also be believed to be fast, so that it is easy to overinvest in this particular service.

### 3.2.3 Feedback with cutoff

There may also be feedback with cutoff such that $\dot{u} = \gamma_{-1}(1-u)/u$ for $u < u_1$ and $\dot{u} = 0$ for $u \geq u_1$.

This implies that potential subscribers lose interest in the product when a certain number of people have bought the product. This can also be regarded as particular segmentation of the market, where the potential subscribers belong to a special interest group. When everyone in the group has bought the product, then no one else will buy it. In this model also, the initial demand is huge for then to become smaller as time increases.

The solution is the same as for the original equation up to $u = u_1$ and constant from this point onward:

$$e^u(1-u) = e^{-\gamma_{-1}t} \text{ for } t < t_1,$$

$$u = u_1 \text{ for } t \geq t_1.$$

From the transcendental equation $-\gamma_{-1}t_1 = u_1 + \ln(1 - u_1)$, $u_1$ is determined, for example, using Newton's method.

### 3.2.4 Feedback tends linearly to zero when everyone has a subscription

The network effect may also be such that the attractiveness is huge when few people owns the product and approaches zero as $u$ approaches one, for example,

$$\dot{u} = \gamma_{11} \frac{(1-u)^2}{u}.$$

In this case, the feedback term is $F(u) = (1-u)/u$. The equation can be written

$$\frac{udu}{(1-u)^2} = -\frac{du}{1-u} + \frac{du}{(1-u)^2} = \gamma_{11}dt$$

The solution satisfying $u = 0$ for $t = 0$ is easily found:

$$(1-u)\exp\left(\frac{u}{1-u}\right) = e^{\gamma_{11}t},$$

or

$$t = \frac{1}{\gamma_{11}}\left[\ln(1-u) + \frac{u}{1-u}\right]$$

where time is expressed as a function of the number of users.

We observe that for $t = \infty$, $u = 1$ since

$$e^{\gamma_{11}t} = (1-u)\exp\left(\frac{u}{1-u}\right) = (1-u)\left[1 + \frac{u}{1-u} + \frac{1}{2!}\left(\frac{u}{1-u}\right)^2 + \frac{1}{3!}\left(\frac{u}{1-u}\right)^3 + \cdots\right] =$$

$$= 1 + \frac{1}{2!}\frac{u^2}{1-u} + \frac{1}{3!}\frac{u^3}{(1-u)^2} + \cdots \to \infty \text{ for } u \to 1.$$

We find $\gamma_{11} = \frac{0.31}{T_{50}}$ and $T_{10} = 0.019 T_{50}$. If $T_{50} = 5$ years, then $T_{10}$ is only 34 days.

## 3.3 Latency time

The latency times and the time it takes the market to increase from 50% to 60% for different feedback strengths are compared in Table 3.3. The table is organized according to increasing latency time.

Two strategic challenges are evident from the table.

- If the market is a trend (that is, the strength of the feedback decreases as the number of users increases), the market will grow very rapid initially followed by a much slower growth as the market matures. In such markets, it will be hard to capture a large number of customers in reasonable time.
- If the strength of the feedback increases as the number of users increases, the initial growth will be slow followed by rapid growth towards a saturated market. In this case, the service or product may be abandoned before the real growth has started. The latency time increases as the feedback strength increases.

Table 3.3 Latency times

| Feedback | Section | Latency time $T_{10}$ | Late evolution $T_{60} - T_{50}$ |
|---|---|---|---|
| $(1-u)/u$, $u_0 = 0$ | 3.2.4 | 33 days | 4 years 6 months |
| $1/u$, $u_0 = 0$ | 3.2.2 | 1 month 20 days | 3 years 3 months |
| $1-u$, $u_0 = 0$ | 3.2.1 | 6 months 18 days | 2 years 6 months |
| No feedback | 2.1 | 9 months | 1 year 7 months |
| $\sqrt{u}$, $u_0 = 0$ | 3.1.3 | 1 year 10 months | 10 months 12 days |
| $u$, $u_0 = 0.01$ | 3.1.2 | 2 years 7 months | 5 months 12 days |
| $u^2$, $u_0 = 0.01$ | 3.1.4 | 4 years 4 months | 25 days |

# 4. MARKETS WITH MORE THAN ONE SUPPLIER

## 4.1 General mathematical model

The general model for the evolution of markets with $n$ suppliers can be expressed as a set of $n$ first order differential equations, one equation for each supplier:

$$\dot{u}_i = M_i(u_1, u_2, \cdots, u_n) + C_i(u_1, u_2, \cdots, u_n),$$

where $M_i(u_1, u_2, \cdots, u_n)$ is the average probability that a new customer will chose supplier $i$ in the time interval $dt$ and $C_i(u_1, u_2, \cdots, u_n)$ is the net number of customers supplier $i$ will receive from or lose to other suppliers during the same interval. The most common market function is that proposed by Bass [Bas]:

$$M_i = \left(1 - \sum_{j=1}^{n} u_j\right)(m_i + r_i u_i),$$

where $m_i$ and $r_i$ are constants. The model is described in Section 3.1 for a market with one supplier.

There are to sub-models. One sub-model contains only innovators; that is, $r_i = 0$. The other sub-model contains only imitators; that is, $m_i = 0$. In the first sub-model, the initial value of the market shares may be zero, that is, $u_i^0 = 0$. In the second sub-model, we must have $u_i^0 > 0$ for at least one $u_i$, since, otherwise, $u_i = 0$ for all $i$ and all time will be a solution to the market equation.

Other methods for evaluating the market evolution are stochastic growths models [Art] and simulation using, for example, Polya urns [Øve]. These methods are not considered here.

## 4.2 Equilibrium points

Since churning does not produce new customers ($\sum C_i = 0$), we get immediately

$$\sum_{i=1}^{n} \dot{u}_i = \sum_{i=1}^{n} (M_i + C_i) = \left(1 - \sum_{j=1}^{n} u_j\right) \sum_{i=1}^{n} (a_i + r_i u_i).$$

The condition $\dot{u}_i = 0$ must be fulfilled for all $i$ at the fixed points. The fixed points are therefore lying on the $(n-1)$-dimensional hyperplane $\sum u_i = 1$. Eventually, the system will end up in one of the fixed points where the coordinates of the point is determined by the market coupling constants $m_i$ and $r_i$, the initial values $u_i^0$, and the form of the churning functions. This fixed point is the attractor for the particular set of parameters and initial values. The coordinates of the attractor will be denoted $(u_1^\infty, u_2^\infty, \cdots, u_n^\infty)$, where the infinity mark indicates that this is the asymptotic value of $u_i$ for $t \to \infty$.

## 4.3 Markets without churning

Let us first consider markets without churning, for example, markets for durable goods such as refrigerators. The differential equations are:

$$\dot{u}_i = \left(1 - \sum_{j=1}^{n} u_j\right)(m_i + r_i u_i).$$

Note that the $u_i$ are non-decreasing functions of time since the suppliers will never lose any of the customers already captured. Dividing each of these equations with the equation for $u_1$ we get

$$\frac{du_i}{du_1} = \frac{m_i + r_i u_i}{m_1 + r_1 u_1}$$

with solution ($r_i > 0$ for all $i$),

$$u_i = \left(\frac{m_i}{r_i} + u_i^0\right)\left(\frac{m_1 + r_1 u_1}{m_1 + r_1 u_1^0}\right)^{r_i/r_1} - \frac{m_i}{r_i}$$

In order to find the location of the fixed points, we must solve the following transcendental equation for $u_1^\infty$:

$$\sum_1^n u_i^\infty = 1 = \sum_1^n \left[\left(\frac{m_i}{r_i} + u_i^0\right)\left(\frac{m_1 + r_1 u_1^\infty}{m_1 + r_1 u_1^0}\right)^{r_i/r_1} - \frac{m_i}{r_i}\right].$$

The value of $u_i^\infty$ for arbitrary $i$ is then obviously:

$$u_i^\infty = \left(\frac{m_i}{r_i} + u_i^0\right)\left(\frac{m_1 + r_1 u_1^\infty}{m_1 + r_1 u_1^0}\right)^{r_i/r_1} - \frac{m_i}{r_i}.$$

It is, therefore, easy to find the coordinates of the attractor. On the other hand, in order to find the temporal evolution of the market, we may have to solve the original set of differential equations using numerical methods, except in the case where all customers are innovators. In this case, the differential equations are

$$\dot{u}_i = \left(1 - \sum_1^n u_j\right) m_i.$$

Summing the equations and setting $U = \sum_1^n u_i$, we find

$$U = 1 - e^{-\sum_1^n m_i t},$$

from which we easily derive the equations for the market share of each supplier

$$u_i = \frac{m_i}{\sum_{j=1}^n m_j}\left(1 - e^{-\sum_{j=1}^n m_j t}\right)$$

with asymptotic solution

$$u_i^\infty = \frac{m_i}{\sum_{j=1}^n m_j}.$$

If all customers are imitators, the differential equations can only be solved using numerical methods. However, the fixed points are easily found to be:

$$u_i^\infty = u_i^0 \left(\frac{u_1^\infty}{u_1^0}\right)^{r_i/r_1}.$$

## 4.4 Market with churning

### 4.4.1 Churning functions

The churning function was discussed in Section 1.4. The general form of the churning function is:

$$C_i = \sum_{j \neq i} a_{ji} u_j f_i(u_i, u_j) - u_i \sum_{j \neq i} a_{ij} f_j(u_j, u_i),$$

where $f_i(u_i, u_j)$ is the market feedback in favor of supplier $i$ (or the popularity of that supplier).

Churning may be *spontaneous* or *stimulated*. Spontaneous churning means that the probability that a customer changes from supplier $i$ to supplier $j$ is independent of the behavior of other customers. Hence, there is no feedback from the market affecting the behavior of spontaneously churning customers. Stimulated churning represents the case where the churning behavior of a customer depends on the behavior of other customers; that is, depends on market feedback. A supplier may experience that both spontaneous and stimulated churning may take place simultaneously.

The condition that there is no net churning is that $C_i = 0$ for all $i$. In this state, churning may still take place but the net result is that each supplier is on average receiving exactly as many customers as the supplier is losing. If there are $n$ suppliers, this gives rise to $n$ equations in the $u_i$. Because of the relation $\sum_i C_i = 0$, not all the equations are independent. In the case of fully developed markets, we may choose any set of $n-1$ equations of the form $C_i = 0$ together with the equation $\sum_1^n u_i = 1$ representing the condition that everyone has become a customer. This gives $n$ independent equations for the number of customers of each supplier.

### 4.4.2 Spontaneous churning

**Asymptotic solution**

If there is no feedback from the market, $f(u_i, u_j) = 1$. The churning function is then

$$C_i = \sum_{j \neq i} a_{ji} u_j - \sum_{j \neq i} a_{ij} u_i = \sum_{j \neq i} a_{ji} u_j - a_{ii} u_i,$$

where $a_{ij} \geq 0$ and $a_{ii} = \sum_{j \neq i} a_{ij}$ are constants. The condition that there is no net churning in the fully developed market (defined as $\sum_1^n u_i = 1$) is that the set of equations $C_i = 0, i = 1 \ldots n-1$ and $\sum_1^n u_i = 1$ are fulfilled simultaneously. This corresponds to the asymptotic solution of the market equations from which we will determine the attractor $(u_1^\infty, u_2^\infty, \cdots, u_n^\infty)$ corresponding to $t \to \infty$. In matrix form, the equations are

$$A \begin{pmatrix} u_1^\infty \\ u_2^\infty \\ \vdots \\ u_{n-1}^\infty \\ u_n^\infty \end{pmatrix} = \begin{pmatrix} -a_{11} & a_{21} & \cdots & a_{n1} \\ a_{12} & -a_{22} & \cdots & a_{n2} \\ \vdots & \vdots & & \vdots \\ a_{1(n-1)} & a_{2(n-1)} & \cdots & a_{n(n-1)} \\ 1 & 1 & \cdots & 1 \end{pmatrix} \begin{pmatrix} u_1^\infty \\ u_2^\infty \\ \vdots \\ u_{n-1}^\infty \\ u_n^\infty \end{pmatrix} = \begin{pmatrix} 0 \\ 0 \\ \vdots \\ 0 \\ 1 \end{pmatrix},$$

with solution

$$\begin{pmatrix} u_1^\infty \\ u_2^\infty \\ \vdots \\ u_{n-1}^\infty \\ u_n^\infty \end{pmatrix} = A^{-1} \begin{pmatrix} 0 \\ 0 \\ \vdots \\ 0 \\ 1 \end{pmatrix} = D^{-1} \begin{pmatrix} A_{11} & A_{21} & \cdots & A_{n1} \\ A_{12} & A_{22} & \cdots & A_{n2} \\ \vdots & \vdots & & \vdots \\ A_{1(n-1)} & A_{2(n-1)} & \cdots & A_{n(n-1)} \\ A_{1n} & A_{2n} & \cdots & A_{nn} \end{pmatrix} \begin{pmatrix} 0 \\ 0 \\ \vdots \\ 0 \\ 1 \end{pmatrix},$$

where the $A_{ij}$ are the cofactors and $D$ is the determinant of $A$. The solution can then be written in the form

$$u_i^\infty = \frac{A_{ni}}{D}.$$

If all the churning coefficients are identical ($a_{ij} = a$), then $u_i^\infty = 1/n$, as expected. This follows directly from the equation

$$\begin{pmatrix} -(n-1)a & a & \cdots & a \\ a & -(n-1)a & \cdots & a \\ \vdots & \vdots & & \vdots \\ a & a & \cdots & a \\ 1 & 1 & \cdots & 1 \end{pmatrix} \begin{pmatrix} u_1^\infty \\ u_2^\infty \\ \vdots \\ u_{n-1}^\infty \\ u_n^\infty \end{pmatrix} = \begin{pmatrix} 0 \\ 0 \\ \vdots \\ 0 \\ 1 \end{pmatrix}$$

by subtracting $a$ times the last row from all the other rows of the matrix. The resulting equation is

$$\begin{pmatrix} -na & 0 & \cdots & 0 \\ 0 & -na & \cdots & 0 \\ \vdots & \vdots & & \vdots \\ 0 & 0 & \cdots & 0 \\ 1 & 1 & \cdots & 1 \end{pmatrix} \begin{pmatrix} u_1^\infty \\ u_2^\infty \\ \vdots \\ u_{n-1}^\infty \\ u_n^\infty \end{pmatrix} = \begin{pmatrix} -a \\ -a \\ \vdots \\ -a \\ 1 \end{pmatrix},$$

from which we derive the desired solution $u_i^\infty = 1/n$.

See, for example, [Ait] for a comprehensive and highly readable introduction to matrix algebra.

We also see immediately that if all $a_{ji} = 0$ for supplier $i$ (the supplier is only losing customers), then this supplier obviously will end up with zero market share (in this case, $A_{ni} = 0$). Similarly, if suppliers 1 through $k$ are only losing customers, then $u_i^\infty = 0, (i \leq k)$. On the other hand, if two or more suppliers are not losing any customers (i.e., $a_{ij} = 0$ for supplier $i$), then matrix $A$ becomes singular and the fixed point cannot be found using this method. In this case we have to solve the dynamic problem and then find the solution for $t \to \infty$.

For the special case of two suppliers we find easily

$$(u_1^\infty, u_2^\infty) = \left( \frac{a_{21}}{a_{12} + a_{21}}, \frac{a_{12}}{a_{12} + a_{21}} \right);$$

that is, the suppliers share the market in accordance with the churning rates.

## Dynamic evolution of the market

The general case using the complete Bass model, results in a set of nonlinear differential equations which cannot be solved analytically. However, if all customers are innovators, the market evolution is determined by a set of linear differential equations which is easily solvable.

The set of differential equations for the market with only innovators is

$$\dot{u}_i = m_i \left( 1 - \sum_j u_j \right) + \sum_{j \neq i} a_{ji} u_j - a_{ii} u_i = m_i - \sum_j q_{ij} u_j,$$

or in vector form

$$\dot{\boldsymbol{u}} = \boldsymbol{m} - Q\boldsymbol{u}$$

where the bold letters are $n$-dimantional column vectors, and $\boldsymbol{m} = (m_1, m_2, \ldots, m_n)^T$ (the symbol **T** denotes transposition). The matrix $Q$ is:

$$Q = (q_{ij}) = \begin{pmatrix} m_1 + a_{11} & m_1 - a_{21} & \cdots & m_1 - a_{n1} \\ m_2 - a_{12} & m_2 + a_{22} & \cdots & m_2 - a_{n2} \\ & & \vdots & \\ m_n - a_{1n} & m_n - a_{2n} & \cdots & m_n + a_{nn} \end{pmatrix}$$

As before, $a_{ii} = \sum_{j \neq i} a_{ij}$. In a realistic market, $m_i > a_{ji}$, so that all the elements of the matrix are positive. Moreover, we consider only the case where the coefficients $m_i$ and $a_{ij}$ are constants.

The solution of the set of differential equations with initial condition $\boldsymbol{u}(0) = 0$ is easily evaluated using the method of integrating factor [Kor], [Goe]:

$$\boldsymbol{u} = (I - e^{-Qt})Q^{-1}\boldsymbol{m} = P(I - e^{-\Lambda t})\Lambda^{-1}P^{-1}\boldsymbol{m}$$

where $I$ is the $n \times n$ unit matrix, $P$ is the matrix composed of the eigenvectors of $Q$, $\Lambda = P^{-1}QP$ is the diagonalization of $Q$, and $Q^{-1} = P\Lambda^{-1}P^{-1}$. We have used the fact that $Pe^{-\Lambda t}P^{-1} = e^{-P\Lambda P^{-1}t} = e^{-Qt}$. The exponential of a matrix is by definition

$$e^{-Qt} = \sum_0^\infty \frac{1}{k!}(-1)^k Q^k t^k$$

containing only elementary operations on the matrix (addition and multiplication), and where $Q^0 = I$. This gives for the exponential of the diagonal matrix $\Lambda$

$$e^{-\Lambda t} = \sum_0^\infty \frac{1}{k!}(-1)^k \begin{pmatrix} \lambda_1 & \cdots & 0 \\ \vdots & \ddots & \vdots \\ 0 & \cdots & \lambda_n \end{pmatrix}^k t^k = \sum_0^\infty \frac{1}{k!}(-1)^k \begin{pmatrix} \lambda_1^k & \cdots & 0 \\ \vdots & \ddots & \vdots \\ 0 & \cdots & \lambda_n^k \end{pmatrix} t^k =$$

$$= \begin{pmatrix} e^{-\lambda_1 t} & \cdots & 0 \\ \vdots & \ddots & \vdots \\ 0 & \cdots & e^{-\lambda_n t} \end{pmatrix}$$

where the $\lambda_i$ are the eigenvalues of $Q$.

The exponential is evaluated by finding the eigenvalues and the corresponding eigenvectors of the matrix $Q$ (see [Kor] or [Gor] for diagonalization algorithms). This requires much computation since the eigenvalues are the roots of algebraic equations of degree $n$. However, the case $n = 2$ is simple.

If all eigenvalues are unique and positive, the solutions in the general case are evidently of the form

$$u_i = \frac{A_{ni}}{D} - \sum_j y_{ij} e^{-\lambda_j t},$$

where the $\lambda_j$ are the eigenvalues of the matrix $Q$ and $y_{ij}$ are functions of the elements of $Q$ and $\boldsymbol{m}$. The equation shows that the asymptotic solution $u_i = A_{ni}/D$ is an attractor; that is, the stable point in which the system eventually ends up for $t \to \infty$.

In the case of two suppliers, we find

$$u_1 = \frac{a_{21}}{a_{12} + a_{21}} - \frac{a_{21}m_2 - a_{12}m_1}{(a_{12} + a_{21})(m_1 + m_2 - a_{12} - a_{21})} e^{-(a_{12}+a_{21})t}$$
$$- \frac{m_1 - a_{21}}{m_1 + m_2 - a_{12} - a_{21}} e^{-(m_1+m_2)t},$$

$$u_2 = \frac{a_{12}}{a_{12} + a_{21}} + \frac{a_{21}m_2 - a_{12}m_1}{(a_{12} + a_{21})(m_1 + m_2 - a_{12} - a_{21})} e^{-(a_{12}+a_{21})t}$$
$$- \frac{m_2 - a_{12}}{m_1 + m_2 - a_{12} - a_{21}} e^{-(m_1+m_2)t},$$

Asymptotically, the solution converges to the attractor $(u_1^\infty, u_2^\infty) = \left(\frac{a_{21}}{a_{12}+a_{21}}, \frac{a_{12}}{a_{12}+a_{21}}\right)$ as already claimed.

If $a_{12} = 0$, that is, supplier 1 is not losing customers, then we find

$$u_1 = 1 - (m_1 - a_{21})e^{-(m_1+m_2)t} - m_2 e^{-a_{21}t},$$

$$u_2 = m_2\left(e^{-a_{21}t} - e^{-(m_1+m_2)t}\right),$$

with attractor $(u_1^\infty, u_2^\infty) = (1,0)$. In this case, the time when supplier 2 has the maximum number of customers is

$$T_m = \frac{\ln(m_1 + m_2) - \ln a_{21}}{m_1 + m_2 - a_{21}}.$$

**Periodic churning coefficients**

In this case, we assume

- that all customers have bought the service; that is, $\sum_1^n u_i = 1$,
- that the churning coefficients are periodic functions of time; that is, $a_{ij} = a_{ij}^0 + \varepsilon_{ij}(t)$ where $\varepsilon_{ij}(t)$ is a periodic function of time with zero mean and $a_{ij}^0$ are constants.[12] The $\varepsilon_{ij}$ may have different periods. For simplicity we assume that these periods are integer multiples of a shortest period.

We then investigate the evolution of the system for $t \to \infty$ starting from some point $(u_1^0, u_2^0, \cdots, u_n^0)$ on the hyperplane $\sum_1^n u_i = 1$. One aim is to find the attractors of the system as $t \to \infty$.

There are $n$ differential equations of the form

$$\dot{u}_i = a_{1i}u_1 + a_{2i}u_2 + \cdots + a_{(i-1)i}u_{i-1} - a_{ii}u_i + a_{(i+1)i}u_{i+1} + \cdots + a_{(n-1)i} + a_{ni}u_n,$$

where only $n - 1$ of them are independent. As before, we set $a_{ii} = \sum_{j \neq i} a_{ij}$. Eliminating $u_n = 1 - \sum_1^{n-1} u_i$, gives a set of $n - 1$ independent differential equations of the form

$$\dot{u}_i = a_{ni} + (a_{1i} - a_{ni})u_1 + \cdots + (a_{(i-1)i} - a_{ni})u_{i-1} - (a_{ii} + a_{ni})u_i + (a_{(i+1)i} - a_{ni})u_{i+1}$$
$$+ \cdots + (a_{(n-1)i} - a_{ni})u_{n-1}$$

or in matrix form,

$$\dot{\boldsymbol{u}} = \boldsymbol{a} - Q\boldsymbol{u}$$

where $\boldsymbol{u} = (u_1, u_2, \cdots, u_{n-1})^\mathrm{T}$, $\boldsymbol{a} = (a_{n1}, a_{n2}, \cdots, a_{n(n-1)})^\mathrm{T}$, and $Q$ is the matrix

---

[12] Periodic functions are introduced to simplify the calculations. We could have chosen stochastic functions with zero mean and small range. However, this would not lead easily to analytic results.

$$Q = \begin{pmatrix} a_{11} + a_{n1} & -a_{21} + a_{n1} & \cdots & -a_{(n-1)1} + a_{n1} \\ -a_{12} + a_{n2} & a_{22} + a_{n2} & \cdots & -a_{(n-1)2} + a_{n2} \\ \vdots & \vdots & \ddots & \vdots \\ -a_{1(n-1)} + a_{n(n-1)} & -a_{2(n-1)} + a_{n(n-1)} & \cdots & a_{(n-1)(n-1)} + a_{n(n-1)} \end{pmatrix}.$$

The $(n-1)$-vector $\boldsymbol{a}$ and the $(n-1) \times (n-1)$-matrix $Q$ are periodic functions of time. Using the method of integrating factor, the solution of the differential equation can be written the form

$$\boldsymbol{u} = \exp\left[-\int_0^t Q(x)dx\right]\left\{\int_0^t \left[\exp\int_0^x Q(y)dy\right]\boldsymbol{a}(x)dx + \boldsymbol{u_0}\right\}$$

where $\boldsymbol{u} = \boldsymbol{u_0}$ is the initial condition for $t = 0$.

We may use Floquet's theorem to investigate this solution [Hop2], [Inc]. Floquet's theorem states that the $(n-1) \times (n-1)$ principal fundamental matrix solution $\Phi(t)$ of the homogeneous differential equation $\dot{\Phi} = -Q\Phi$ is

$$\Phi(t) = \exp\left[\int_0^t Q(x)dx\right] = S(t)e^{Rt}, \Phi(0) = I,$$

where $I$ is the unit matrix, $S(t)$ is a periodic matrix and $R$ is a constant matrix. The columns of $\Phi(t)$ are $n-1$ linearly independent solutions of the differential equation for $\boldsymbol{u}$. The solution for $\boldsymbol{u}$ is then

$$\boldsymbol{u}(t) = \int_0^t S(t)^{-1}S(x)e^{-R(t-x)}\boldsymbol{a}(x)dx + S(t)^{-1}e^{-Rt}\boldsymbol{u_0}$$

Since

$$\int f(t)e^{\alpha t}dt = g(t)e^{\alpha t} + h(t)$$

where $f$, $g$ and $h$ are periodic functions of $t$, the solution for $u_i$ may be expressed as the sum of three terms:

$$u_i(t) = A_i + B_i(t) + \sum_j C_{ij}(t)e^{-\lambda_j t}$$

where $A_i$ are constants, $B_i(t)$ and $C_{ij}(t)$ are periodic functions of time, and the $\lambda_j$ are the eigenvalues of $R$. It is difficult to find the matrices $S(t)$ and $R$.

In the case of two suppliers, it is easy to find an analytic solution. The differential equation is

$$\dot{u}_1 = a_{21}^0 + \varepsilon_{21}(t) - \left(a_{21}^0 + \varepsilon_{21}(t) + a_{12}^0 + \varepsilon_{12}(t)\right)u_1$$

since $u_1 + u_2 = 1$.

The solution is easily found using the method of integrating factor

$$u_1(t) = \frac{a_{21}^0}{a_{12}^0 + a_{21}^0} + P(t) + B(t)e^{-(a_{12}^0 + a_{21}^0)t},$$

where

$$P(t) = \frac{1}{1+\alpha(t)}\left[-\frac{\alpha(t)a_{21}^0}{a_{12}^0 + a_{21}^0} + \sigma(t)\right],$$

$$B(t) = \frac{1}{1+\alpha(t)}\left[u_1^0 - \frac{a_{21}^0}{a_{12}^0 + a_{21}^0}\right]$$

$$\alpha(t) = \exp\left[\int_0^t [\varepsilon_{12}(x) + \varepsilon_{21}(x)]dx\right] - 1,$$

$$\sigma(t) = \int_0^t [a_{21}^0\alpha(x) + \epsilon_{21}(x) + \epsilon_{21}(x)\alpha(x)]\exp[-(a_{12}^0 + a_{21}^0)(t-x)]dx,$$

where $u_1^0$ is the initial value of $u_1$.

Hence, the solution then consists of three parts as predicted for the general case:

- The market share without oscillations $a_{21}^0/(a_{12}^0 + a_{12}^0)$.
- A periodic function $P(t)$ with zero mean. This is easily seen by expanding the functions $\varepsilon_{ij}$ in Fourier series and then integrating.
- An asymptotically vanishing function $B(t)e^{-(a_{12}^0+a_{21}^0)t}$ which represents the motion from the chosen initial value $u_1^0$ to the average market share.

### 4.4.3 Stimulated churning

#### Only stimulated churning

In this case, $f_i(u_i, u_j) = u_i g_i(u_i, u_j)$ for all suppliers. The equation expresses the condition that the number of customers supplier $i$ receives from supplier $j$ is proportional to the number of customers of supplier $i$. The condition that no net churning takes place is then:

$$\sum_{j\neq i} a_{ji}u_iu_jg_i(u_i, u_j) = u_i \sum_{j\neq i} a_{ij}u_jg_j(u_j, u_i), \quad \sum_j u_j = 1,$$

or

$$\sum_{j\neq i} u_iu_j[a_{ji}g_i(u_i, u_j) - a_{ij}g_j(u_j, u_i)] = 0, \quad \sum_j u_j = 1.$$

The state where $a_{ji}g_i(u_i, u_j) = a_{ij}g_j(u_j, u_i)$ cannot correspond to stable fixed points since any infinitesimal change in the $a_{ij}$ and the $g_i$ will force the system to move away from this point. Therefore, all except one of the $u_i$ must be zero at a stable fixed point. Together with the condition $\sum_i u_i = 1$ we then conclude that any of the points $(1,0,0,\cdots,0)$, $(0,1,0,\cdots,0)$, $(0,0,1,\cdots,0)$, $\cdots$, $(0,0,0,\cdots,1)$ may be fixed points of the market equation. However, only one of these points can be an attractor. Which of the points is the attractor in a given market depends on the form of the churning function, the $a_{ij}$, and the initial value of the $u_i$. These markets may then be called winner-take-all markets.

Note that, in this case, the equations for the temporal evolution of the market are nonlinear differential equations, and the simple methods of Section 4.4.2 cannot be used to determine the dynamic solutions. See, for example, [But].

## Both spontaneous and stimulated churning

Another simple case is where stimulated churning takes place only toward one of the suppliers, say supplier 1, while all suppliers are subject to spontaneous churning; that is, $f_1(u_1, u_j) = b_1 u_1 + 1$ $(j \neq 1)$, $f_i(u_i, u_j) = 1$ $(i \neq 1, j \neq i)$. The equilibrium conditions are

$$C_1 = \sum_{j \neq 1}(b_{j1} u_1 u_j + a_{j1} u_j - a_{1j} u_1) = 0, (b_{j1} = b_1 a_{j1}),$$

$$C_i = \sum_{j \neq i}(a_{ji} u_j - a_{ij} u_i) - b_{i1} u_1 u_i = 0, (i \neq 1),$$

$$\sum_j u_j = 1.$$

If $a_{1j} > 0$, the first equation has a solution only if supplier 1 and at least one of the other suppliers have nonzero market share at asymptotic equilibrium. Hence, the asymptotic market must be shared by at least to suppliers.

However, if supplier 1 is not losing customers because of churning (that is, if $a_{1j} = 0$), then the attractor is the point $(1, 0, \cdots, 0)$, or supplier 1 captures the entire market.

If there are only two suppliers, we have two independent equations

$$C_1 = b_{21} u_1 u_2 + a_{21} u_2 - a_{12} u_1 = 0,$$

$$u_1 + u_2 = 1$$

with solutions

$$2 b_{21} u_1 = b_{21} - a_{12} - a_{21} + \sqrt{(b_{21} + a_{12} + a_{21})^2 - 4 a_{21} b_{21}},$$

$$u_2 = 1 - u_1.$$

We see that the solution is consistent since it satisfies the inequality $0 < u_1 \leq 1$ for all values of the parameters $b_{21}$, $a_{12}$ and $a_{21}$.

If $a_{12} = 0$, the solution is $u_1 = 1$ and $u_2 = 0$ as just claimed.

More generally, the market feedback may be written as $f_i(u_i, u_j) = b_i u_i + \varepsilon_i$ where $\varepsilon_i$ is either 0 or 1. At the attractor, the equation $\sum_j u_j = 1$ and $n - 1$ equations of the form

$$\sum_{j \neq i}[(b_i a_{ji} - b_j a_{ij}) u_i u_j + \varepsilon_i a_{ji} u_j - \varepsilon_j a_{ij} u_i] = 0$$

must be fulfilled. We see that the attractor can be any point on the hyperplane $\sum_j u_j = 1$ depending upon the values of the market parameters.

In the case of two suppliers and $\varepsilon_1 = \varepsilon_2 = 1$, the solution is

$$2(b_{21} - b_{12}) u_1 = b_{21} - b_{12} - a_{12} - a_{21} + \sqrt{(b_{21} - b_{12} + a_{12} + a_{21})^2 - 4 a_{21}(b_{21} - b_{12})}$$

and $u_2 = 1 - u_1$, where $b_{ij} = b_i a_{ji}$.

# 5. GAMES AND SERVICES WITH LIMITED POPULARITY

## 5.1 Introduction

These models are applicable to games and to services where the users lose interest in the service after some time. For simplicity, we will refer to this case as games, however, with the understanding that the models may be valid for certain services also.

The game models described below are based on the following assumption concerning games:

- a person buys at most one copy of a particular game;
- the number of people buying the game is so big that we can describe the dynamics of the market by treating the dependent variables as continuous functions of time;
- in many cases, the market for a particular game is independent of other games;
- there are cases where the market for a particular game depends on the number of people who have bought another game (complementary games);
- there are games where the number of people buying the game is independent of how many have already bought the game (no externalities);
- there are games where the popularity of the game depends on the number of people having bought the game (positive feedback from the market);
- there are games where the likelihood for leaving the game depends on the number of people having left the game.

In all models, there are three actors: those who may buy a given game ($B$ – potential buyers), the actual players using the game actively ($P$ – players), and those who have quitted the game (or never bought it) ($Q$ – quitters). The model is called the BPQ market model. $N$ designates either the whole population or the part of the population that may buy the game.

## 5.2 The BPQ market model

### 5.2.1 General model

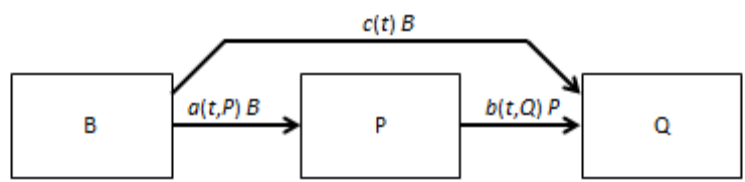

Figure 5.1 Market model

The market model is shown in Figure 5.1. The flow parameters are:

- $a(t,P)$ is the intensity of new players buying the game. The parameter may depend on both time and the number of players already playing the game; that is, there may be a positive feedback from the market or a network effect encouraging new players to enter the game;

- $b(t, Q)$ is the intensity of players quitting the game. This parameter may depend on the number of players having quitted the game. This is also a positive feedback from the network [Can];
- $c(t)$ is the intensity of players that will never buy the game; this parameter reflects that the interest for the game in the population may vary with time. This parameter is only used in the model where $a(t, P)$ is independent of $P$ in order to investigate the effect of this parameter on $P$.

We assume that the population potentially interested in the game is constant and equal to $N$. The flow parameters may be constants or functions of time.

We will develop models for several forms of the parameters $a(t, P)$ and $b(t, Q)$. The models are listed in the order of how difficult it is to find analytic solutions of the corresponding differential equations.

Case 1: $a(t, P) = a(t)$ and $b(t, Q) = b(t)$. Both the general case and the special case where $a(t), b(t)$, and $c(t)$ are constants will be considered. Using the terminology of Boss, all potential players are innovators. In this case we can derive analytic solutions to the model. The effect of the parameter $c(t)$ is investigated.

Case 2: $a(t, P) = \beta P$, $b(t, Q) = b$ and $c(t) = 0$. The parameters $\beta$ and $b$ are constants. This model is identical to the SIR model of epidemiology, and we can import solutions of that theory directly into our market model. In Boss terminology, all players are imitators.

Case 3: $a(t, P) = a + \beta P$, $b(t, Q) = b$ and $c(t) = 0$. In this case the parameters $a, \beta$ and $b$ are constants. In this model, the differential equations can only be solved using numerical methods. In this case, the players are a mixture of innovators and imitators.

Case 4: $a(t, P) = \beta P$, $b(t, Q) = \gamma Q$ and $c(t) = 0$. The parameters $\beta$ and $\gamma$ are constants. This is the case where there are two feedback loops from the market: one is stimulating new players to enter the game as in case 2, and the other stimulates players to quit the game depending upon how many who have quitted the game.

Case 5: $a(t, P) = a$, $b(t, Q) = \gamma Q$ and $c(t) = 0$, where $a$ and $\gamma$ are constants. In this case, players leaving the game stimulate others to do the same.

Case 6: $a(t, P) = a$, $b(t, Q) = b + \gamma Q$ and $c(t) = 0$, where $a, b$ and $\gamma$ are constants.

It follows directly from Figure 5.1 that the following set of differential equations describes the evolution of the market:

$$\dot{B}(t) = -\bigl(a(t, P) + c(t)\bigr)B(t),$$

$$\dot{P}(t) = a(t, P)B(t) - b(t, Q)P(t),$$

$$\dot{Q}(t) = b(t, Q)P(t) + c(t)B(t).$$

We observe immediately that $\dot{B}(t) + \dot{P}(t) + \dot{Q}(t) = 0$, from which we derive the obvious conservation law $B(t) + P(t) + Q(t) = N$ for all $t$. Moreover, $B(t)$ is a monotonically decreasing function of time and that $Q(t)$ is a monotonically increasing function of time. $P(t)$ has a maximum at time $T_m$ given by the condition

$$\dot{P}(T_m) = 0 \rightarrow a(T_m, P(T_m))B(T_m) = b(T_m, Q(T_m))P(T_m).$$

If the coupled set of differential equations cannot be solved analytically, they can be solved numerically for all functions $a(t, P)$, $b(t, Q)$, and $c(t)$ using, for example, the Runge-Kutta method for coupled first order differential equations (See [But]).

The demand, $D(t)$, defined as the number of players entering the game per unit time, and the number of people having bought the game, $C(t)$, at time $t$, are, respectively.

$$D(t) = a(t, P(t))B(t)$$

$$C(t) = \int_0^t D(u)du = \int_0^t a(u, P(u))B(u)du.$$

### 5.2.2 Case 1: $a(t, P) = a(t)$ and $b(t, Q) = b(t)$

**General solution**

The differential equations for $B$, $P$ and $Q$ are now

$$\dot{B}(t) = -(a(t) + c(t))B(t),$$

$$\dot{P}(t) = a(t)B(t) - b(t)P(t),$$

$$\dot{Q}(t) = b(t)P(t) + c(t)B(t)$$

Natural initial conditions are $B(0) = N$ and $P(0) = Q(0) = 0$.
The first equation gives immediately,

$$B(t) = N \exp\left\{-\int_0^t [a(u) + c(u)]du\right\}.$$

The differential equation for $P$ is then

$$\dot{P}(t) = Na(t) \exp\left\{-\int_0^t [a(u) + c(u)]du\right\} - b(t)P(t).$$

This is a non-homogeneous first order linear differential equation with solution

$$P(t) = N \exp\left\{-\int_0^t b(u)du\right\} \int_0^t a(u) \exp\left\{-\int_0^u [a(w) + c(w) - b(w)]dw\right\} du.$$

We find $Q$ from $Q(t) = N - B(t) - P(t)$. Observe that the boundary conditions $B(0) = N$, $P(0) = 0$, and $Q(0) = 0$ are automatically fulfilled since $\int_0^0 f(u)du = 0$ for all well-behaved functions $f(t)$.

From the equation for $\dot{P}(t)$, we find that the gradient $\dot{P}(0)$ at $t = 0$ is $\dot{P}(0) = Na(0)$. This is the rate at which the market increases just after the game has been launched.

Note that the initial gradient depends only on the parameter $a(t)$, so that for small $t$, any game of this type can be approximated as $P(t) = Na(0)t$.

The demand and the number of people having bought the game at time $t$ are

$$D(t) = a(t)N \exp\left\{-\int_0^t [a(u) + c(u) - b(u)]du\right\}$$

$$C(t) = \int_0^t D(u)du.$$

**Special case where $a, b,$ and $c$ are constants**

Let us now consider the special case where $a, b,$ and $c$ are constants.

In this case we find for $b \neq a + c$

$$B(t) = Ne^{-(a+c)t},$$

$$P(t) = \frac{aN}{a+c-b}\left(e^{-bt} - e^{-(a+c)t}\right),$$

$$Q(t) = \frac{N}{a+c-b}\left[a+c-b-ae^{-bt} + (b-c)e^{-(a+c)t}\right].$$

The solution for $P(t)$ for small $t$ is $P(t) = aNt[1 - (a+b+c)t/2]$ to the second order in $t$. The function is concave. This expression may be used to estimate initial values for the parameters $a, b,$ and $c$ for further curve fitting purposes.

The demand and the number of players are

$$D(t) = aNe^{-(a+c)t},$$

$$C(t) = \frac{aN}{a+c}\left(1 - e^{-(a+c)t}\right).$$

The total number of players is then

$$C(\infty) = \frac{aN}{a+c}.$$

The maximum of $P(t)$ is located at

$$T_m = \frac{\ln(a+c) - \ln b}{a+c-b}$$

giving

$$P_m = \frac{Na}{a+c-b}\left(e^{-bT_m} - e^{-(a+c)T_m}\right).$$

The parameters $a$ and $b$ can be estimated by choosing $T_m$ (say, 1 year) and the ratio $(a+c)/b$ (say, 2), and then compute $a+c$. For $T_m = 1$ year and $(a+c)/b = 2$, we get $a + c = \ln 4 = 1.4$ year$^{-1}$.

## Confluence

The above formulas do not apply in the confluent case where $b = a + c$. In this case, we find easily $P(t) = Nate^{-bt}$.

The maximum of $P(t)$ is then found at

$$T_m = \frac{1}{b},$$

giving

$$P_m = N\frac{a}{b}e^{-1}.$$

Moreover.

$$D(t) = aNe^{-bt},$$

$$C(t) = \frac{aN}{b}(1 - e^{-bt}).$$

## Special case where $a(t) = a_0 + a_1 t$, and $b$ and $c$ are constants

We find:

$$P = Ne^{-bt}\int_0^t (a_0 + a_1 u) \exp\left\{-\left[(c-b)u + a_0 u + \frac{a_1 u^2}{2}\right]\right\} du =$$

$$= Ne^{-bt}\int_0^t (a_0 + a_1 u)e^{K^2}\exp\left[-\left(u\sqrt{\frac{a_1}{2}} + K\right)^2\right] du = NIe^{-bt+K^2},$$

where

$$K = \frac{(a_0 + c - b)}{\sqrt{2a_1}}$$

and

$$I = \int_0^t (a_0 + a_1 u) \exp\left[-\left(u\sqrt{\frac{a_1}{2}} + K\right)^2\right] du.$$

Setting $q = \sqrt{\frac{a_1}{2}}$ and using $v = uq + K$ as new independent variable, we get

$$I = q^{-1}\int_K^{K+qt} (a_0 - 2qK + 2qv)\, e^{-v^2} dv =$$

$$= \sqrt{\frac{\pi}{2a_1}}(a_0 - 2qK)\left[\text{erf}(K + qt) - \text{erf}(K)\right] - 2q[-e^{-(K+qt)^2} + e^{-K^2}],$$

where

$$\text{erf}(x) = \frac{2}{\sqrt{\pi}} \int_0^x e^{-u^2} du.$$

Finally,

$$P = Ne^{-bt+K^2} \left\{ \sqrt{\frac{\pi}{2a_1}} (a_0 - 2qK)[\text{erf}(K+qt) - \text{erf}(K)] - 2q[e^{-K^2} - e^{-(K+qt)^2}] \right\}$$

where $K = (a_0 + c - b)/\sqrt{2a_1}$ and $q = \sqrt{a_1/2}$.

More generally, $a(t)$ may be any piecewise smooth function of time. In this case, the particular function was chosen so that the integrals could be solved analytically.

**Special case where $b(t) = b_0 + b_1 t$, and $a$ and $c$ are constants**

We find:

$$P = N \exp\left[-\left(b_0 t + \frac{b_1 t^2}{2}\right)\right] \int_0^t a \exp\left[-(a+c-b_0)u + \frac{b_1 u^2}{2}\right] du =$$

$$= NIa \exp\left[-\left(b_0 t + \frac{b_1 t^2}{2}\right)\right],$$

where

$$I = \int_0^t \exp\left[-(a+c-b_0)u + \frac{b_1 u^2}{2}\right] du.$$

We find

$$I = re^{-G^2} \int_{-G}^{rt-G} e^{u^2} du,$$

where $G = (a+c-b_0)\sqrt{2b_1}$ and $r = \sqrt{b_1/2}$. The integral must be evaluated using numerical methods.

The total solution is

$$P = Nar\, e^{-(G^2 + b_0 t + \frac{b_1 t^2}{2})} \int_{-G}^{rt-G} e^{u^2} du.$$

### 5.2.3 Case 2: $a(t, P) = \beta P$, $b(t, Q) = b$ and $c = 0$

**Equivalence to the SIR model of epidemiology**

The model is shown in the figure. Note that, in the terminology of Bass, there are no innovators but only imitators participating in the game.

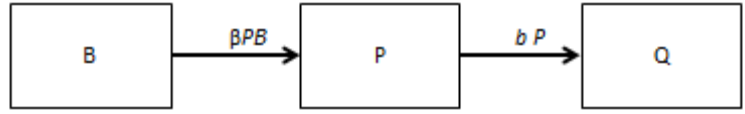

Figure 5.2 SIR model

The differential equations are:

$$\dot{B}(t) = -\beta P(t) B(t),$$

$$\dot{P}(t) = \beta P(t) B(t) - bP(t),$$

$$\dot{Q}(t) = bP(t),$$

where $\beta$ and $b$ are constants. Moreover, $B(t) + P(t) + Q(t) = N$ for all $t$. The condition that $P(t)$ is a maximum is simply $B(T_m) = b/\beta$.

This model is identical to the SIR model of epidemiology with the transformation $S \to B, I \to P, R \to Q$, where $S$ is the number susceptible, $I$ is the number infectious, and $R$ is the number recovered [Mur]. We can then immediately write down some important results.

**Initiation problem**

We see directly that $B(t) = N, P(t) = 0, Q(t) = 0$ is a solution of the differential equations. This means that if there are no players initially, there will be no players in the future. Therefore, we must have $P(0) = P_0 > 0$. Moreover, we may set $B(0) = B_0 = N - P_0$ and $Q(0) = 0$. If $P(t) \ll N$, we have approximately $B(t) = N$ for small $t$. The second differential equation is then for small $t$

$$\dot{P}(t) = \beta P(t) N - bP(t),$$

with solution

$$P(t) = P_0 e^{(\beta N - b)t} = P_0[1 + (\beta N - b)t + (\beta N - b)^2 t^2/2] + \mathcal{O}(t^3)$$

for small $t$. This function is convex. The initial growth is slow if $P_0 \ll N$.

This observation leads to the following strategic dilemmas: When the game is launched, the provider must stimulate a number of users to start playing the game; otherwise, no one will play it.

Even if there are initial players, it still may take a long time before a significant number of player take part in the game. We may also say that the latency time of the game is long. Therefore, the game may be abandoned before it really takes off. A similar phenomenon was studied in Section 3 for ordinary markets with positive feedback.

**Relationships between $B$, $P$, and $Q$**

Simple relationships between the functions $B(t), P(t)$ and $Q(t)$ are derived easily from the differential equations.

Dividing the first equation with the last equation gives

$$\frac{dB(t)}{dQ(t)} = -B(t)$$

from which it follows that

$$B(t) = B_0 \exp\left[-\frac{b}{\beta} Q(t)\right],$$

$$P(t) = N - B(t) + \frac{b}{\beta} \ln \frac{B(t)}{B_0},$$

$$Q(t) = N - B(t) - P(t),$$

$$B(\infty) = B_0 \exp\left[-\frac{\beta}{b}(N - B(\infty))\right].$$

The last equation gives $B(\infty)$ as solution of a transcendental equation from which we can compute the total number of players $P_{tot} = Q(\infty) = N - B(\infty)$, since eventually all players have quitted the game. Moreover, the maximum of $P(t)$ is determined by $\dot{P}(t) = 0$; that is,

$$B(T_m) = \frac{b}{\beta},$$

$$Q(T_m) = \frac{b}{\beta} \ln \frac{\beta B_0}{b},$$

$$P(T_m) = N - \frac{b}{\beta} - \frac{b}{\beta} \ln \frac{\beta B_0}{b}.$$

The differential equation for $Q(t)$ is

$$\dot{Q}(t) = bP(t) = b\left\{N - Q(t) - B_0 \exp\left[-\frac{\beta}{b} Q(t)\right]\right\}.$$

This gives

$$t = \frac{1}{b} \int_0^Q \frac{du}{N - u - B_0 e^{-\frac{\beta u}{b}}}.$$

This allows us to draw $Q$ as a function of time. The corresponding value for $B$ is computed from the equation expressing $B$ as a function of $Q$, and from $P = N - Q - B$, we finally find $P$. Hence, all the variables may then be computed as a function of time by solving the integral.

We also see that $T_m$ is then:

$$T_m = \frac{1}{b} \int_0^{(b/\beta)\ln(\beta B_0/b)} \frac{du}{N - u - B_0 e^{-\frac{\beta u}{b}}}.$$

Knowing $P(T_m)$, we find $b/\beta$ by solving the nonlinear equation for $P(T_m)$ numerically. Also knowing $T_m$, we find $b$ from the equation above.

### 5.2.4 Case 3: $a(t, P) = a + \beta P(t)$, $b(t, Q) = b$ and $c = 0$

The differential equations are:

$$\dot{B}(t) = -(a + \beta P(t))B(t),$$

$$\dot{P}(t) = (a + \beta P(t))B(t) - bP(t),$$

$$\dot{Q}(t) = bP(t),$$

where $a$, $\beta$ and $b$ are constants. As before, $B(t) + P(t) + Q(t) = N$ for all $t$.

The simplest approach is to solve the following set of coupled differential equations numerically:

$$\dot{B}(t) = -(a + \beta P(t))B(t),$$

$$\dot{P}(t) = (a + \beta P(t))B(t) - bP(t),$$

and observe that $Q(t) = N - B(t) - P(t)$. The differential equations are easily solved numerically using the Runge-Kutte method for coupled differential equations.

The initial growth of the market is given by $P(t) \xrightarrow[t \to 0]{} aNt[1 + (N\beta - b)t/2]$. This formula is derived by setting $B(t) = N$ in the differential equation for $P(t)$ and then solving it by simple quadrature, keeping only terms up to time squared. This function is convex. This function may then be used as a first estimate of the parameters $a, b,$ and $\beta$ by comparing the theoretical growth curve and the observed initial growth curve of the game, and using this estimate as initial values for better curve fitting.

Dividing the second equation with the first, we find:

$$\frac{dP}{dB} = -1 + \frac{bP}{(a + \beta P)B}.$$

It is easy to solve this equation numerically so that we may draw $P$ as a function of $B$. We also see that the maximum $P(T_m)$ of $P$ satisfies

$$P(T_m) = \frac{aB(T_m)}{b - \beta B(T_m)},$$

where $B(T_m)$ is the corresponding value of $B$.

### 5.2.5 Case 4: $a(t, P) = \beta P$, $b(t, Q) = \gamma Q$ and $c = 0$[13]

The differential equations are:

$$\dot{B}(t) = -\beta P(t)B(t),$$

$$\dot{P}(t) = \beta P(t)B(t) - \gamma P(t)Q(t),$$

$$\dot{Q}(t) = \gamma P(t)Q(t).$$

The interpretation of the last equation is that the likelihood that a player quits the game is not only proportional to the number of players but also to the number of players having quitted the game. Hence, the act that a player leaves the game is therefore

---
[13] See [Can]

encouraging other players to leave the game. This is again a positive feedback from the market.

$P(t) = 0, B(t) = N, Q(t) = 0$ is obviously a solution of the differential equations, so that we must have $P(0) = P_0 > 0$ for non-trivial solutions. Similarly, if $Q(0) = Q_0 = 0$, then the $Q(t) = 0$ for all time, so that the whole population will eventually become players of the game; that is, $P(t) \xrightarrow{t \to \infty} N$. This is also an uninteresting solution. Therefore, we have to assume that $Q_0 > 0$, or, in other words, there must be an initial population of quitters. This may be interpreted as a group of peoples who is not interested in the game, and who inspires other people not to enter the game. We also see from the third equation that $\dot{Q}_0 = \gamma P_0 Q_0$. This result will be used later.

We find immediately, dividing the first differential equation with the third equation and integrating:

$$B(t) = B_0 \left(\frac{Q_0}{Q(t)}\right)^{\beta/\gamma} = (N - P_0 - Q_0)\left(\frac{Q_0}{Q(t)}\right)^{\beta/\gamma}.$$

We see again that $Q_0$ must be larger than zero; otherwise, the equation for $B(t)$ has no solution.

Inserting $P = \dot{Q}/\gamma Q$ and computing the derivative $\dot{P}$, the second equation can be written as a second order differential equation in $Q$:

$$\ddot{Q} = \frac{\dot{Q}^2}{Q} + \beta B_0 \dot{Q} \left(\frac{Q_0}{Q}\right)^{\beta/\gamma} - \gamma \dot{Q} Q.$$

If we insert

$$\ddot{Q} = \frac{d\dot{Q}}{dt} = \frac{d\dot{Q}}{dQ}\frac{dQ}{dt} = \frac{d\dot{Q}}{dQ}\dot{Q},$$

the second order differential equation in $t$ is reduced to a first order differential equation for $\dot{Q}$ as a function of $Q$,

$$\frac{d\dot{Q}}{dQ} = \frac{\dot{Q}}{Q} + \beta B_0 \left(\frac{Q_0}{Q}\right)^{\beta/\gamma} - \gamma Q.$$

This equation is linear in $\dot{Q}$, and the solution (also called the first integral) is easily found:

$$\dot{Q} = \gamma Q \left[-B_0 \left(\frac{Q_0}{Q}\right)^{\beta/\gamma} - Q + C\right],$$

where $C$ is a constant of integration found by setting $\dot{Q}_0 = \gamma P_0 Q_0 = \gamma Q_0(-B_0 - Q_0 + C)$. This gives $C = N$. The solution where $t$ is given as a function of $Q$ is then:

$$t = \int_{Q_0}^{Q} \frac{du}{\gamma u \left[N - B_0 \left(\frac{Q_0}{u}\right)^{\beta/\gamma} - u\right]}.$$

Together with

$$B(t) = B_0 \left(\frac{Q_0}{Q(t)}\right)^{\beta/\gamma}$$

and
$$P(t) = N - B(t) - Q(t)$$
we have a complete solution of the equations.

From the original differential equation, we see that $P(t)$ has a maximum at $T_m$ if $\beta B(T_m) = \gamma Q(T_m)$. Moreover, $B(T_m) = B_0[Q_0/Q(T_m)]^{\beta/\gamma}$. This gives

$$Q(T_m) = \left[\frac{\beta B_0 Q_0^{\beta/\gamma}}{\gamma}\right]^{\frac{\gamma}{\beta+\gamma}},$$

and, finally

$$P(T_m) = N - \left(1 + \frac{\gamma}{\beta}\right)\left[\frac{\beta B_0 Q_0^{\beta/\gamma}}{\gamma}\right]^{\frac{\gamma}{\beta+\gamma}}.$$

### 5.2.6 Case 5: $a(t,P) = a$, $b(t,Q) = \gamma Q$ and $c(t) = 0$

In this model, there is no market externality encouraging users to enter the play. On the other hand, there is an externality encouraging players to leave the game. The differential equations are:

$$\dot{B}(t) = -aB(t),$$

$$\dot{P}(t) = aB(t) - \gamma P(t)Q(t),$$

$$\dot{Q}(t) = \gamma P(t)Q(t).$$

We see that $Q(t) = 0$ is a solution of the third equation leading to trivial solutions for $P(t)$ and $B(t)$, so that we must have $Q_0 > 0$ for the existence of nontrivial solutions. Setting $P_0 = 0$, we find $B_0 = N - Q_0$. The solution of the first equation is then $B(t) = B_0 e^{-at}$.

Inserting $Q(t) = N - P(t) - B(t)$ in the equation for $P$ gives

$$\dot{P}(t) = aB(t) - \gamma P(t)\big(N - P(t) - B(t)\big) = \gamma P(t)^2 - \gamma P(t)(N - B_0 e^{-at}) + aB_0 e^{-at}.$$

This is a Riccati equation. We see that $N - B_0 e^{-at}$ is a particular solution, corresponding to the solution $Q(t) = 0$ for all $t$. The Riccati equation can then be transformed to a linear differential equation by the transformation (see [Kor], [Inc])

$$P(t) = N - B_0 e^{-at} - \frac{1}{u(t)},$$

giving

$$\dot{u}(t) = -\gamma(N - B_0 e^{-at})u(t) + \gamma$$

The solution is

$$u(t) = \left[\gamma \int_0^t \exp\left(\gamma Nx + \frac{\gamma B_0}{a}e^{-ax}\right)dx + \frac{1}{Q_0}e^{-\gamma B_0/a}\right]\exp\left[-\left(\gamma Nt + \frac{\gamma B_0}{a}e^{-at}\right)\right],$$

satisfying the initial condition $u(0) = (N - B_0 - P_0)^{-1} = 1/Q_0$. $P(t)$ and $Q(t)$ are easily computed from this equation.

The integral has to be evaluated using numerical integration.

The condition that $P(t)$ has a maximum is $aB(T_m) = \gamma P(T_m)Q(T_m)$. Moreover. $B(T_m) = B_0 e^{-aT_m}$ and $Q(T_m) = N - P(T_m) - B(T_m)$. Combining these equations, we find that the maximum satisfies a quadratic equation with solution

$$2P(T_m) = N - B_0 e^{-aT_m} + \sqrt{(N - B_0 e^{-aT_m})^2 - \frac{4a}{\gamma} B_0 e^{-aT_m}}.$$

Choosing $T_m$ and $P(T_m)$, the equation my used to estimate values for $a$ and $\gamma$.

### 5.2.7 Case 6: $a(t,P) = a$, $b(t,Q) = b + \gamma Q$ and $c(t) = 0$

In this model, some players are stimulated to quit the game and some leaves the game without being stimulated to do so.

$$\dot{B}(t) = -aB(t),$$

$$\dot{P}(t) = aB(t) - [b + \gamma Q(t)]P(t),$$

$$\dot{Q}(t) = [b + \gamma Q(t)]P(t).$$

As in the previous case, the solution for $B(t)$ is readily found to be $B(t) = Ne^{-at}$. $B(0) = N$, $P(0) = 0$, and $Q(0) = 0$ can now be chosen as initial conditions. Inserting $B(t) = Ne^{-at}$ and $Q(t) = N - P(t) - B(t)$ in the second equation, we find that $P(t)$ satisfies the Riccati equation

$$\dot{P}(t) = \gamma P(t)^2 - (b + \gamma N - \gamma Ne^{-at})P(t) + aNe^{-at}.$$

There is no simple particular solution to this equation so that the equation cannot be solves using the same method as in case 5. However, the equation is easily solved using numerical methods.

## 5.3 Complementary games

### 5.3.1 Complementary games without market feedback

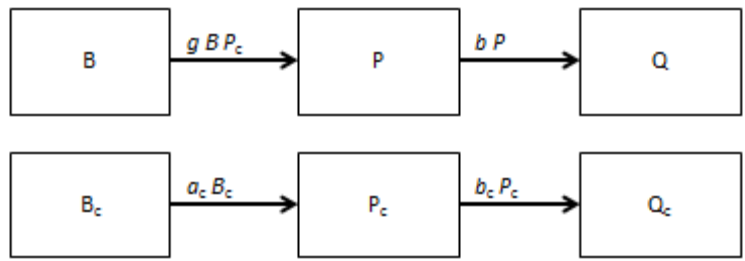

Figure 5.3 Game$_1$ depends on the popularity of game$_2$

In this model, $B(t)$ and $B_C(t)$ are the number of people having bought game 1 or 2, respectively, at time $t$, where $B(0) = N$. We are concerned with the case where $B_C(0)$ is different from 0; that is the complementary game 2 was marketed before game 1. The term $gBP_c$ reflects that the number of sold games of type 1 is proportional to the number of complementary games of type 2 being sold. The differential equations are

$$\dot{B} = -gBP_c$$

$$\dot{P} = gBP_c - bP$$

$$\dot{Q} = bP$$

$$\dot{B}_c = -a_c B_c$$

$$\dot{P}_c = a_c B_c - b_c P_c$$

$$\dot{Q}_c = b_c P_c$$

where $g, a_c, b$ and $b_c$ are constants.

The initial conditions for game 1 are $B(0) = N, P(0) = Q(0) = 0$. The solution for $P_c$ is the same as the solution for $P$ in Section 5.2.2 with $c = 0$. However, the complementary game may have been marketed earlier than game 1; say, at time $t = -\tau$, while game 1 is introduced at time $t = 0$. Note that if $\tau = 0$, then the two games are marketed at the same time, and if $\tau$ is negative, game 2 was marketed later than game 1. The solution for game 2 is then

$$P_c = \frac{Na_c}{b_c - a_c}\left[e^{-a_c(t+\tau)} - e^{-b_c(t+\tau)}\right].$$

We can now treat $gP_c$ as a time-dependent parameter $\alpha(t)$; that is,

$$\dot{B} = -\alpha(t)B,$$

$$\dot{P} = \alpha(t)B - bP.$$

We get:

$$B = N \exp\left[-\int_0^t \alpha(u)du\right],$$

$$P = Ne^{-bt}\int_0^t \alpha(u)\exp\left[\int(-\alpha(u) + b)du\right]du = Ne^{-bt}\int_0^t \frac{dA(u)}{du}e^{-A(u)}e^{bu}du,$$

where

$$\alpha(u) = \frac{Nga_c}{b_c - a_c}\left[e^{-a_c(u+\tau)} - e^{-b_c(u+\tau)}\right],$$

$$A(u) = \int \alpha(u)du = \frac{Ng}{b_c(b_c - a_c)}\left[a_c e^{-b_c(u+\tau)} - b_c e^{-a_c(u+\tau)}\right].$$

Integrating by parts, the solution is

$$P = N(e^{-A(0)-bt} - e^{-A(t)}) + Nb \int_0^t e^{-A(u)+b(u-t)} du.$$

The integral is evaluated using numerical integration.

### 5.3.2 Complementary games with positive market feedback

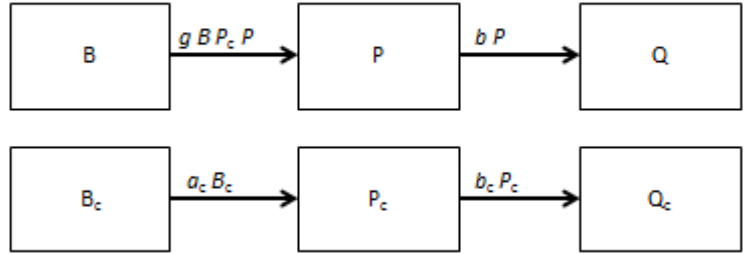

Figure 5.4 Game$_1$ depends on the popularity of both games

The equations are now:

$$\dot{B} = -gBP_c$$

$$\dot{P} = gBP_c - bP$$

$$\dot{Q} = bP$$

$$\dot{B}_c = -a_c B_c$$

$$\dot{P}_c = a_c B_c - b_c$$

$$\dot{Q}_c = b_c P_c$$

where $h, b, a_c$ and $b_c$ are constants.

The solution for $P_c$ is as before:

$$P_c = \frac{Na_c}{b_c - a_c}\left[e^{-a_c(t+\tau)} - e^{-b_c(t+\tau)}\right].$$

However, the equations for $B, P$ and $Q$ cannot be solved with simple methods since now $gP_c$ is a function of time. Therefore, the solution method of Section 5.2.3 is no longer applicable. On the other hand, we may estimate the effect of the complimentary game by simply setting the number of items sold equal to a constant value $P_{c0}$. We may then set $gP_c = gP_{c0} = g_c$, and we are back to the case in Section 5.2.3. The effect of the complementary game will then be to alter (increase) the number of players adopting game 1 per unit of time.

# References

[Art] W. Brian Arthur, Yu. M. Ermoliev, and Yu. M. Kaniovski, Strong Laws for a Class of Path-Dependent Stochastic Processes with Applications, in *Proc. International Conf. on Stochastic Optimization, Kiev 1984,* Lecture Notes in Control and Information Science, IIASA 81, Springer, 1986.

[Ait] A. C. Aitken, *Determinants and Matrices (University Mathematical Texts)*, Praeger, 1983

[Can] John Cannarella and Joshua A. Spechler, Epidemiological modeling of online social network dynamics, arXiv.org:1401.4208

[Bas] Frank M. Bass, A New Product Growth Model for Consumer Durables, *Management Science*, Vol. 15, No. 5, 1969

[But] John C. Butcher, *Numerical methods for ordinary differential equations* (2nd ed.), John Wiley & Sons, 2008

[Goe] Gerald Goertzel and Nunzio Tralli, *Some Mathematical Methods of Physics*, McGraw-Hill, 1960

[Hop] F. C. Hoppensteadt and C. S. Peskin, *Modeling and Simulation in Medicine and the Life Sciences* ($2^{nd}$ edition), Springer, 2002

[Hop2] F. C. Hoppensteadt, *Analysis and Simulation of Chaotic Systems*, Springer, 1993

[Inc] E. L. Ince, *Ordinary Differential Equations*, Dover, 1956 (invaluable sourcebook on ordinary differential equation; first published in 1926 and still in print)

[Kor] G. A. Korn and T. M. Korn, *Mathematical Handbook for Scientists and Engineers: Definitions, Theorems, and Formulas for Reference and Review*, Dover, 1968

[Mur] D. Murray, *Mathematical Biology,* I. *An Introduction* ($3^{rd}$ edition), Springer, 2002

[Sch] M. Schroeder, *Fractals, Chaos, Power Laws: Minutes from an Infinite Paradise*, Dover, 2009

[Sha] Shapiro, C. and Varian, H. R., *Information Rules: A Strategic Guide to the Network Economy*, 1999, Harvard Business School Press

[Ste] John D. Sterman, , *Business dynamics: Systems thinking and modeling for a complex world*. McGraw Hill, 2000

[Str] Steven H. Strogatz, *Nonlinear Dynamics and Chaos*, Perseus Books Publishing, 1994

[Tes] L. Tesfatsion and K. L. Judd (Eds.), *Handbook of Computational Economics*, Volume 2, *Agent-Based Computational Economics,* Elsevier/North-Holland, 2006

[Øve] H. Øverby, G. Biezók and J. A. Audestad, Modeling Dynamic ICT Services Markets, *2012 World Telecommunications Congress (WTC)*, Miyazaki, Japan, 5-6 March, 2012